**Shape-induced obstacle attraction and repulsion during dynamic locomotion**

Yuanfeng Han, Ratan Othayoth, Yulong Wang, Chun-Cheng Hsu, Rafael de la Tijera Obert, Evains Francois, Chen Li*

Department of Mechanical Engineering, Johns Hopkins University

*Corresponding author. Email: [redacted]

## NOVELTY STATEMENT

Our study expanded the new concept and usefulness of terradynamic shapes for passive control of dynamic robot locomotion via body-obstacle physical interaction, by discovering obstacle attraction and repulsion of a cuboidal body and an elliptical body, respectively. Despite the complex physical interaction, a quasi-static potential energy landscape model explained the observed dependence of turning and pitching motions on body shape and revealed the physical principles without solving equations of motion.

## ABSTRACT

Robots still struggle to dynamically traverse complex 3-D terrain with many large obstacles, an ability required for many critical applications. Body-obstacle interaction is often inevitable and induces perturbation and uncertainty in motion that challenges closed-form dynamic modeling. Here, inspired by recent discovery of a terradynamic streamlined shape, we studied how two body shapes interacting with obstacles affect turning and pitching motions of an open-loop multi-legged robot and cockroaches during dynamic locomotion. With a common cuboidal body, the robot was attracted towards obstacles, resulting in pitching up and flipping-over. By contrast, with an elliptical body, the robot was repelled by obstacles and readily traversed. The animal displayed qualitatively similar turning and pitching motions induced by these two body shapes. However, unlike the cuboidal robot, the cuboidal animal was capable of escaping obstacle attraction and subsequent high pitching and flipping over, which inspired us to develop an empirical pitch-and-turn strategy for cuboidal robots. Considering the similarity of our self-propelled body-obstacle interaction with part-feeder interaction in robotic part manipulation, we developed a quasi-static potential energy landscape model to explain the dependence of dynamic locomotion on body shape. Our experimental and modeling results also demonstrated that obstacle attraction or repulsion is an inherent property of locomotor body shape and insensitive to obstacle geometry and size. Our study expanded the concept and usefulness of terradynamic shapes for passive control of robot locomotion to traverse large obstacles using physical interaction. Our study is also a step in establishing an energy landscape approach to locomotor transitions.

**KEYWORDS**

Obstacle traversal, contact, physical interaction, potential energy landscape, robot, cockroach, terradynamics

## 1. INTRODUCTION

Mobile robots are on the verge of becoming a major part of society. Many important applications, such as search and rescue, environmental monitoring, and extraterrestrial exploration, require robots to move through terrain with many obstacles comparable to or larger that robot's size. Robot locomotion in such complex 3-D environments commonly relies on obstacle avoidance (Borenstein and Koren, 1991; Khatib, 1986; Rimon and Koditschek, 1992). The idea is to use sensors to obtain a geometric map of the environment (Dissanayake et al., 2001; Elfes, 1989), estimate the robot's location and pose (Dissanayake et al., 2001; Leonard and Durrant-Whyte, 1991), treat large objects other than relatively flat ground (for terrestrial locomotion) as obstacles, plan a collision-free trajectory around them (Borenstein and Koren, 1991; Khatib, 1986; Lavalle, 1998), and control (Latombe, 2012) the robot to follow the trajectory. This approach is successful in structured (e.g., vacuum robots in organized rooms) (Tribelhorn and Dodds, 2007) and even dynamic (e.g., self-driving cars in traffic) (Thrun et al., 2000) environments, where obstacles are sparse, high-fidelity sensing and online computation is practical, and locomotor-ground interaction is well understood and readily controlled (Bekker, 1956; Pacejka, 2005; Wong, 2008).

However, obstacle avoidance using envrionment geometry fails for dynamic locomotion in highly cluttered, complex 3-D terrain, where many of the above steps fail. First, a collision-free trajectory to reach the goal simply may not exist. In addition, many existing methods of localization, state estimation, planning, and control of robot locomotion only work well for slow, quasi-static, kinematic movement (Arslan and U. Saranli, 2012). By contrast, inevitable, continual dynamic interaction of the robot with cluttered large obstacles (Gart et al., 2018; Gart and Li, 2018; Li, Pullin, et al., 2015; Qian et al., 2019; Rieser et al., 2019; Transeth et al., 2008) induces frequent large perturbations and uncertainty in motion and sensing (Garcia Bermudez et al., 2012; Ordonez et al., 2013) or even results in flipping-over (Guizzo and Ackerman, 2015; Li et al., 2016), making these methods impractical. Furthermore, exisiting locomotor-ground interaction models (e.g., tyre dynamics for paved roads (Pacejka, 2005), terramechanics for deformable off-road terrain (Bekker, 1956; Wong, 2008)) do not apply to cluttered terrain.

Biologically inspired multi-legged robots hold the promise for robustly traversing cluttered terrain, with inherent advantage of legs in overcoming large obstacles (Raibert, 2008) and higher stability offered by multiple legs (Full et al., 2006; Ting et al., 1994). By exploiting natural dynamics of leg-ground interaction (Childers et al., 2016; De and Koditschek, 2018; Garcia Bermudez et al., 2012; Li et al., 2009; Qian and Koditschek, 2020; Raibert, 2008; Saranli et al., 2001; Spagna et al., 2007) and body-terrain interaction (Gart et al., 2018; Gart and Li, 2018; Li, Pullin, et al., 2015), multi-legged robots have achieved improved locomotor performance and capabilities in complex terrain compared to those using quasi-static planning and control (although still far from robust (Arslan and U. Saranli, 2012)). For example, a recent study empirically discovered that an ellipsoidal body shape helps traverse cluttered obstacles by inducing body rolling through gaps narrower than body width (Li, Pullin, et al., 2015).

Inspired by the critical role of shape to align parts in robotic manipulation (Peshkin and Sanderson, 1988), here we take the next step in understanding how shape affects body-obstacle interaction during dynamic locomotion. We experimentally tested and compared how two body shapes—cuboidal and elliptical—affect pitching and yawing motions (Sections 2), useful for initiating climbing and turning. While the cuboidal body shape (Fig. 2A, top) is typical of multi-legged robots (Fig. 1), an elliptical body shape (Fig. 2A, bottom) is rarely used by robots. For simplicity, we chose rigid vertical pillars as our model terrain (Fig. 2), representative of large unmovable obstacles common in cluttered terrain. To further test the effect of obstacle geometry, we used pillars of circular and square cross-sectional shapes and varied their orientation. To understand to what extent traversal can be achieved using passive interaction and when active adjustments must be made, we comparatively studied a RHex-class robot with open-loop control and insects capable of sensory feedback (Fig. 2). We chose the forest-floor-dwelling discoid cockroach (Fig. 2B) that is exceptional at traversing cluttered obstacles like dense vegetation and leaf litter (Bell et al., 2007). We studied the animal's locomotion during rapid escape response where passive mechanics dominates (Dickinson et al., 2000; Full and Koditschek, 1999; Sponberg and Full, 2008) to allow closer comparison with the open-loop robot.

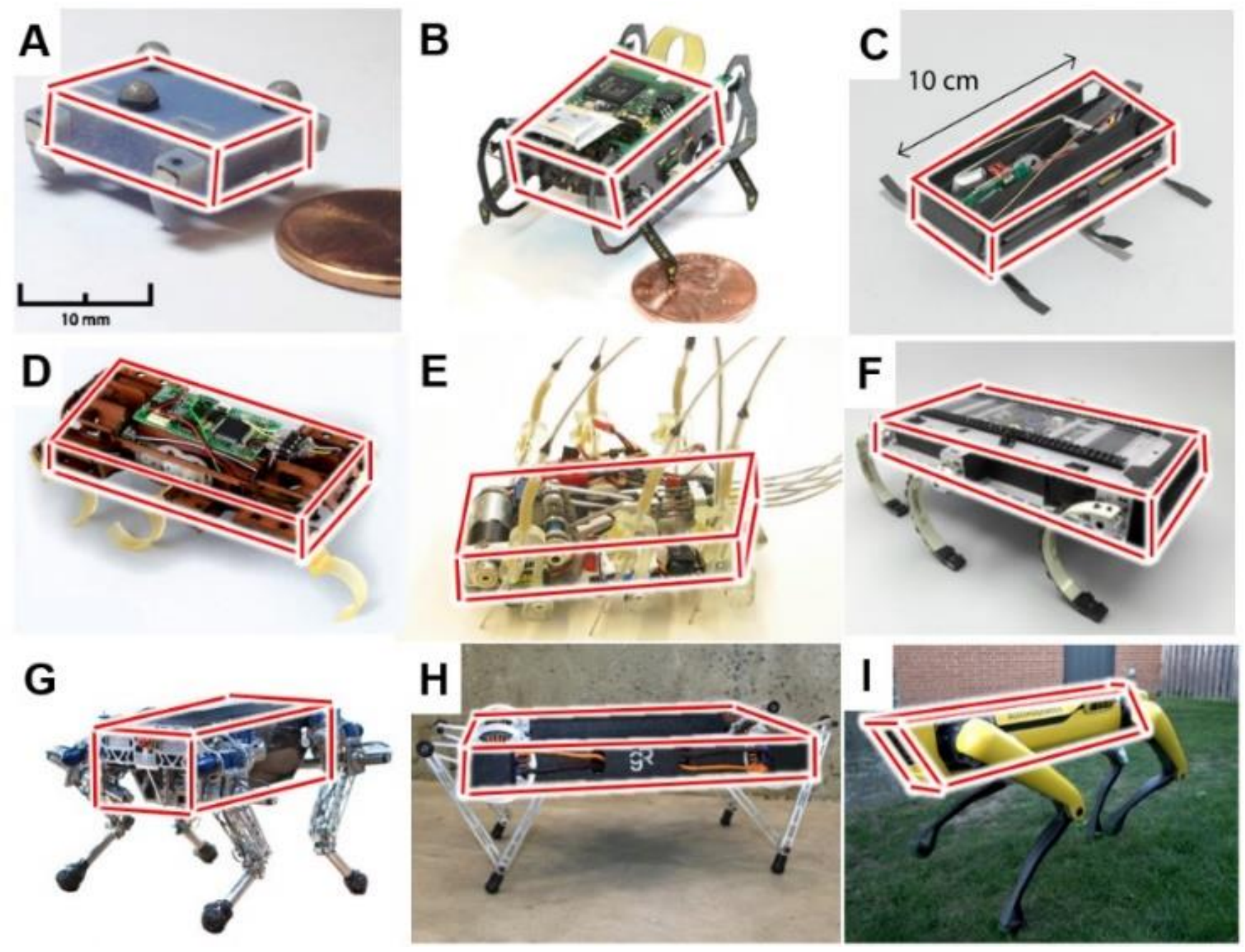

**Fig. 1. Cuboidal body shape typical of multi-legged robots.** (**A**) Mini quadrupedal robot (Pierre and Bergbreiter, 2016). (**B**) Ambulatory Microrobot (Baisch and Wood, 2011). (**C**) DASH (Birkmeyer et al., 2009). (**D**) VelociRoACH (Haldane and Fearing, 2015). (**E**) iSprawl (Kim et al., 2006). (**F**) RHex (Saranli et al., 2001). (**G**) StarlETH (Hutter et al., 2012). (**H**) Minitaur (Blackman et al., 2016). (**I**) SpotMini (Ackerman, 2016).

For both the robot and animal, we discovered that a cuboidal body shape induces attraction towards the obstacle whereas an elliptical body shape induces repulsion away from the obstacle; however, both are surprisingly insensitive to obstacle geometry (Section 3). Inspired by the similarity of physical interaction and usefulness of quasi-static potential energy models for explaining shape-dependent physical interaction in robotic part alignment and grasping without force closure (Brost, 1992; Jayaraman, 1996; Zumel, 1997), we developed a similar potential energy landscape model to explain the sensitive dependence of dynamic interactions on body shape and further generalize our finding to more body shapes and obstacle sizes (Sections 4, 5). The model well explained the observed dependence of physical interaction on body shape and the resulting stochastic dynamic locomotion. In addition, it revealed that interaction is insensitive to obstacle geometry and size and that attractive interaction stems from the body's frontal flatness. We demonstrated the usefulness of shape-induced obstacle repulsion and attraction for passive control of dynamic locomotion in complex terrain by eliciting desired locomotor transitions (Section 6). Finally, we discuss broader implications for robotics and future directions (Section 7).

We note that our work is an advancement over the previous study (Li, Pullin, et al., 2015) in several aspects. First, this work expands the concept of terradynamic shapes, from an ellipsoidal body shape that induces body rolling into obstacle gaps, to cuboidal and elliptical body shapes that induce obstacle attraction and repulsion. These different physical interactions are useful for eliciting distinct locomotor transitions. In addition, the previous study did not study how body-obstacle interaction depends on obstacle shape, orientation, and size. Here, we discover that obstacle attraction or repulsion is an inherent property of the locomotor body shape and is insensitive to obstacles. Furthermore, although the previous study introduced an early potential energy landscape model, in order to explain the observed dependence of body orientation on body shape during obstacle interaction, it assumed that the system is always in its minimal potential energy state over rotational degrees of freedom. However, it is not clear whether this is true for a self-propelled, dynamically moving robot or animal. By quantifying 3-D kinematics and reconstructing body-obstacle interaction, our study reconstructs how the system behaved on the potential energy landscape. We demonstrate that, although the self-propelled system does not reach minimal potential energy orientation, it is strongly attracted to it, explaining why the quasi-static potential energy landscape is useful in understanding dynamic locomotion.

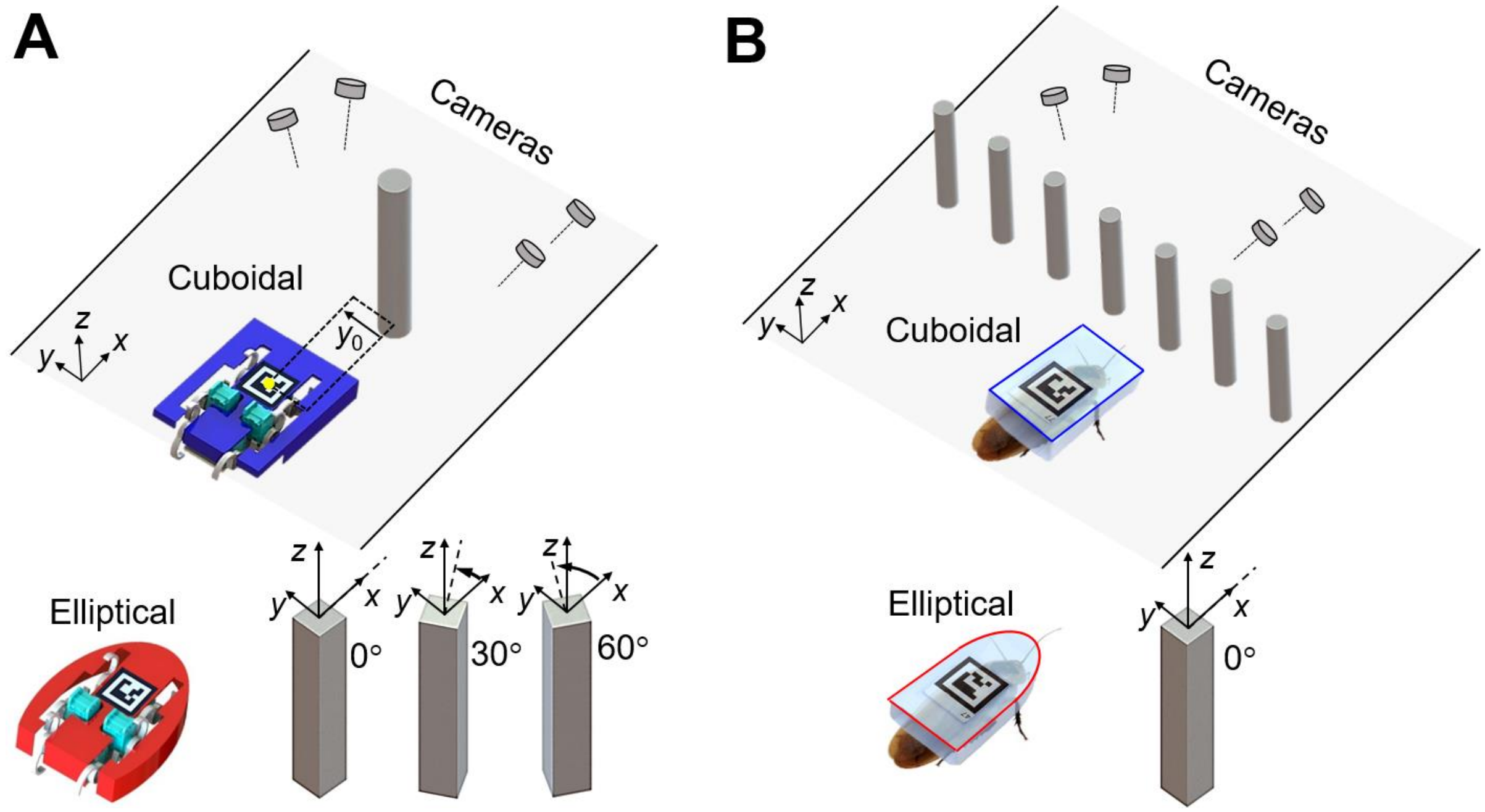

**Fig. 2. Experimental design.** (**A, B**) In robot (A) and animal (B) experiments, an open-loop, sensor-less, six-legged robot and discoid cockroaches, each with a cuboidal or elliptical body (achieved by wearing a shell), interact with pillar(s) of a circular or square cross-sectional shape. See Table S1 for robot and animal body and shell masses and dimensions. $y_0$ is the initial lateral offset of robot from pillar. Schematic of square pillars orientated at 0°, 30°, and 60° relative to $x$-axis are also shown.

# 2. EXPERIMENTAL METHODS

## 2.1. Robot experiments

We built a RHex-class robot for experiments (Saranli et al., 2001). We attached six servo motors with embedded velocity PD controllers to the chassis to drive six 3-D printed S-shaped legs (PLA plastic) in an alternating tripod gait (leg rotation frequency = 1 Hz). We chose S-shaped legs to double effective step frequency over the original C-shaped legs of RHex. To increase ground traction, we wrapped legs with duct tape and covered the robot obstacle track with sandpaper (60 grit size). We attached a 3-D printed elliptical or cuboidal shell to its dorsal surface (Fig. 2A). Because we could precisely control robot trajectory when approaching obstacles, we simply used a single pillar to study robot-terrain interaction. We parameterized square pillar orientation using the angle between its left/right side and the forward ($+x$) direction and chose it to be 0°, 30°, and 60° (Fig. 2A). We started the robot at a constant initial fore-aft distance (50 cm between the robot marker and pillar center) moving at a speed of 30 cm $s^{-1}$, with an initial direction aligned with $x$-axis. We varied initial lateral offset $y_0$ (0 cm, 3 cm, 6 cm, 9 cm, and 12 cm) to vary initial bearing (Fig. 2A). For each pillar shape and orientation, we collected 25 trials, with 5 trials for each initial bearing. See Table S1 in Supplementary Information for sample size, mass, and dimensions of robot experiments.

## 2.2. Animal experiments

We used ten healthy male discoid cockroaches (*Blaberus discoidalis*), as females were often gravid and under different load bearing conditions. Prior to experiments, we kept the animals in individual plastic containers at room temperature (22 °C) on a 12h:12h light:dark cycle and provided dog food and water *ad libitum*. We attached an elliptical or a cuboidal shell to the animal's dorsal surface (Fig. 2B) using hot glue after trimming the wings. The shells were vacuum formed from 0.05 cm thick polystyrene sheets. Because the animal rarely collided if only a single pillar was present, we constructed an obstacle field with a row of seven rigid aluminum pillars which were press-fit vertically onto an acrylic sheet (Fig. 2B). The circular pillars (Fig. 2B, top) had a diameter that equals the side length of square pillars (Fig. 2B, bottom). The lateral spacing between two adjacent pillars was only 10% larger than the animal shell width. We did not vary square pillar orientation as in robot experiments, as natural animal motion already varied it. We tested 10 animals for each body shape and used the same group of animals for different pillar shapes and orientations. For each trial, we placed the animal at one end of the track and pressed its abdomen to induce it to traverse the pillars. The animal was then allowed to rest in a shelter at the opposite end of the track for at least one minute. Animal experiments were conducted at 40 °C. See Supplementary Information, Table S2 for sample size, mass, and dimensions in animal experiments. We verified that during a rapid escape response the animal's body-pillar initial obstacle interaction was dominated by passive mechanics with minimal active sensory feedback (Supplementary Information Section S3, Fig. S3).

### 2.3. 3-D kinematics reconstruction

We recorded robot and animal experiments at 100 frames $s^{-1}$ and a resolution of $2560 \times 2048$ pixels using four synchronized, calibrated cameras from top views with different orientations (Fig. 2). To automatically track the body in all camera views, we attached a BEEtag marker (Crall et al., 2015) to the dorsal face of the shell. We then reconstructed 3-D kinematics of marker using Direct Linear Transformation (Hedrick, 2008) and calculated the body's 3-D position ($x$, $y$, $z$) relative to the pillar and 3-D orientation (Euler angles: yaw $\alpha$, pitch $\beta$, and roll $\gamma$, $Z$-$Y'$-$X''$ Tait-Bryan angle convention[1]). We identified the phases of the trial when body was in contact with the obstacle, by creating a CAD model of the shell and reconstructing its motion relative to the pillar(s) using 3-D kinematics calculated above. See Supplementary Information, Section S1 for details of imaging and reconstruction.

[1] Note that for this convention pitch is negative when the body pitches head up. However, we chose to report body pitch to be positive for head pitching up to be consistent with common definition of pitching up.

### 2.4. Data analysis

We focused on the physical interaction at the first pillar contacted by the robot or animal. To simplify quantification of continual body-obstacle interaction composed of many small high-frequency collisions (Supplementary Information, Section S4), we defined the interaction phase as the portion of the trial during which the body had sustained continual collisions. This begins when the body's front face (curved vertical face for elliptical body) came within a small distance (animal: 1.5 mm, robot: 2 mm) of the pillar and ends when this distance exceeded 5 mm or when the body collided with a second obstacle (whichever occurred sooner). We compared recorded and 3-D reconstructed videos to verify that this definition well matched the visually observed beginning and ending of the interaction in all trials, and we found them to be accurate to within 10 ms. All following analyses of body-obstacle interaction are for this continual interaction phase.

To quantify turning and pitching during interaction, we defined the following metrics. We defined body bearing ($\theta$) relative to the pillar as the angle between the body midline and the line connecting body geometric center and pillar center in the *x-y* plane (Fig. 4A). Initial bearing $\theta_0$ was the value when the body first contacted the pillar. Final bearing $\theta_f$ was the value when the body maximally turned away or towards the pillar relative to its initial bearing (Fig. 4C, D, filled blue and red dots). Similarly, we defined initial pitch ($\varphi_0$) as the body pitch angle ($\varphi$) when the body first contacted the pillar (Fig. 5A) and maximal pitch increase $\Delta\varphi_{max}$ (Fig. 5C, D, blue and red arrows) as the difference between the maximal and initial values.

For the animal, we rejected trials in which the animal did not contact any pillar or contacted the sidewall before contacting any pillar. We also rejected trials in which animal interacted with the first pillar for less than 0.03 s before contacting a second pillar because, when interaction was too short, the initial dynamic collision and body oscillation due to leg-ground interaction dominated motion. With these criteria, we accepted ~5 trials for each animal with each body shape and pillar shape and orientation treatment. We defined transition probability for both the robot and animal as the percentage of trials in which a locomotor mode or outcome occurs among all accepted trials. Note that it is not the same as transition probability of a Markov chain. See Tables S1 and S2 for sample size of accepted trials.

See Supplementary Information, Section S2 for statistical tests.

## 3. EXPERIMENTAL RESULTS

### 3.1. Traversal probability depends on body but not obstacle shape

With a cuboidal body, both the robot and animal had a low traversal probability. The robot never traversed the circular pillar (Fig. 3A), and the animal only traversed about once every three times (Fig. 3C). By contrast, with an elliptical body, both the robot and animal had a high traversal probability (Fig. 3B, D). Surprisingly, traversal probability of both the robot and animal did not sigificantly change with obstacle shape and orientiation (Fig. S4, $P < 0.05$, Kruskal-Wallis test). The difference in traversal probability between the two body shapes was mainly a result of the sensitive dependence of body turning and pitching on body shape, as elaborated below.

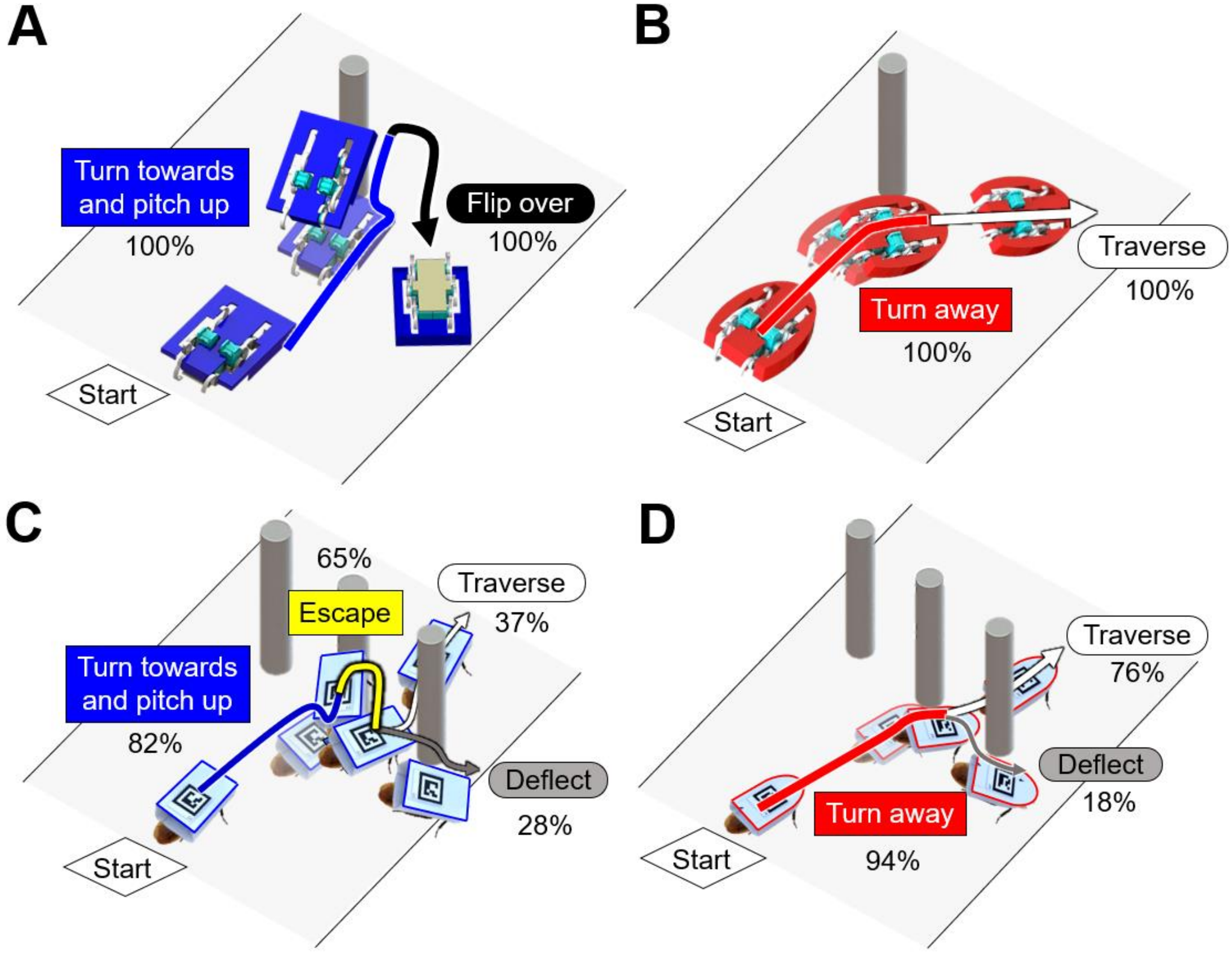

**Fig. 3. Locomotor modes and interaction outcomes are sensitive to body shape.** **(A-D)** Definition of locomotor modes (rectangular boxes) and obstacle interaction outcomes (rounded boxes) and their probabilities[2] for robot (A, B) and animal (C, D) interacting with a circular pillar. Blue and red represent cuboidal and elliptical body shapes, respectively. See Movie S1 for representative trials.

---

[2] In the animal panels, the probabilities of modes and outcomes shown do not sum to 100%. This was because modes that were not our focus were not shown. In these trials (e.g., 18% of all trials in C, 6% in D), after initial contact, the animal either contacted the sidewalls, substantially slowed down using active sensory feedback, or stopped moving.

## 3.2. Turning depends sensitively on body shape

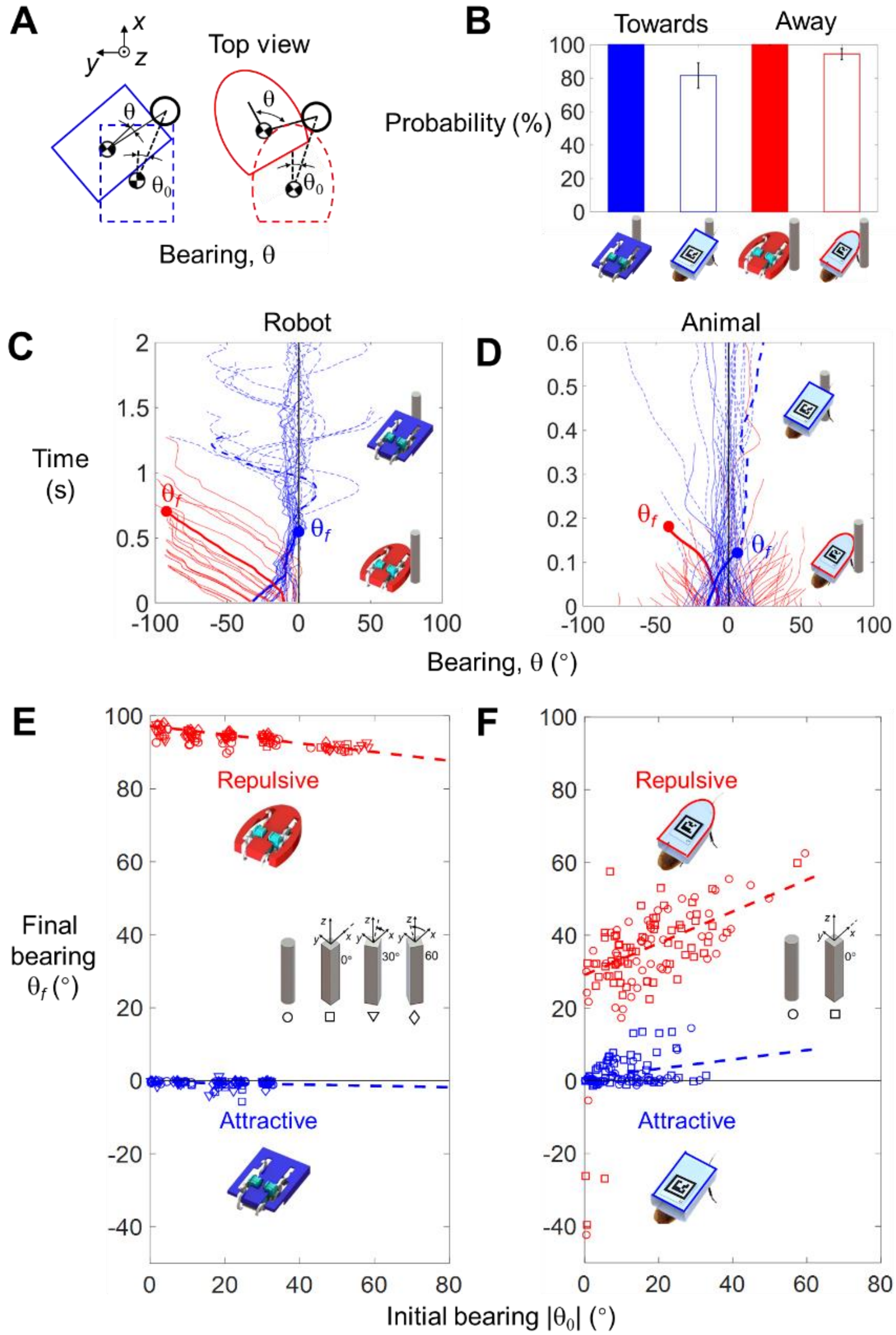

**Fig. 4. Turning motion during obstacle interaction.** (**A**) Definition of body bearing relative to pillar θ. When θ = 0°, body midline is aligned with obstacle. Dashed and solid outlines show body at initial contact

(initial bearing = $\theta_0$) and a later time. (**B**) Probability of robot (filled) and animal (open) being attracted towards obstacle (blue) and repelled away from obstacle (red). (**C, D**) $\theta$ of robot (C) and animal (D) as a function of time for all trials. Thick curve is a representative trial for each treatment, with dot showing final bearing $\theta_f$. In (B-D), data are shown for circular pillars and are similar for other pillar shapes and orientations (Fig. S5). (**E, F**) Final bearing $\theta_f$ of robot (E) and animal (F) as a function of initial bearing magnitude $|\theta_0|$. We flipped data in the third and fourth quadrants to the second and first quadrants considering lateral symmetry. Circle, square, triangle, and diamond symbols are for circular pillars and square pillars oriented at 0°, 30°, and 60° relative to *x*-axis, respectively. Dashed lines are linear fits of data (excluding outliers of $\theta_f < 0$ for elliptical animal data in (F)). In all panels, blue and red represent cuboidal and elliptical body shapes, respectively. See Movie S1 and Movie S2 for representative trials.

The cuboidal and elliptical body shapes resulted in distinct body turning directions (Fig. 4). With a cuboidal body, the robot or animal was initially attracted to and turned towards the obstacle (Fig. 3A, C, blue curve; Fig. 4B, blue; Movie S1), with final bearing $\theta_f$ (see definition in Fig. 4C, D) almost always converging to zero for all $\theta$ (Fig. 4C-F, blue) . Although the robot's bearing often overshot after aligning with the obstacle ($\theta = 0$), it was always attracted back and oscillated around the obstacle (Fig. 4C, blue; Movie S1). The animal, however, rarely oscillated after aligning with the obstacle but often continued to turn away until it traversed the obstacle (Fig. 3C, yellow; Fig. 4D, blue; Movie S1), with continual leg pushing and slipping and occasional body rolling. By contrast, with an elliptical body, the robot or animal was quickly repelled and turned away from the obstacle (Fig. 3B, D, red curve; Fig. 4B, red; Movie S1), with $\theta_f$ almost converging to 90° for the robot for all $\theta_0$ and increasing with $|\theta_0|$ for the animal (Fig. 4C-F, red). After turning away from the first pillar, the animal was sometimes deflected by other pillars (Fig. 3C, D, gray arrow) and did not traverse.

### 3.3. Pitching depends sensitively on body shape

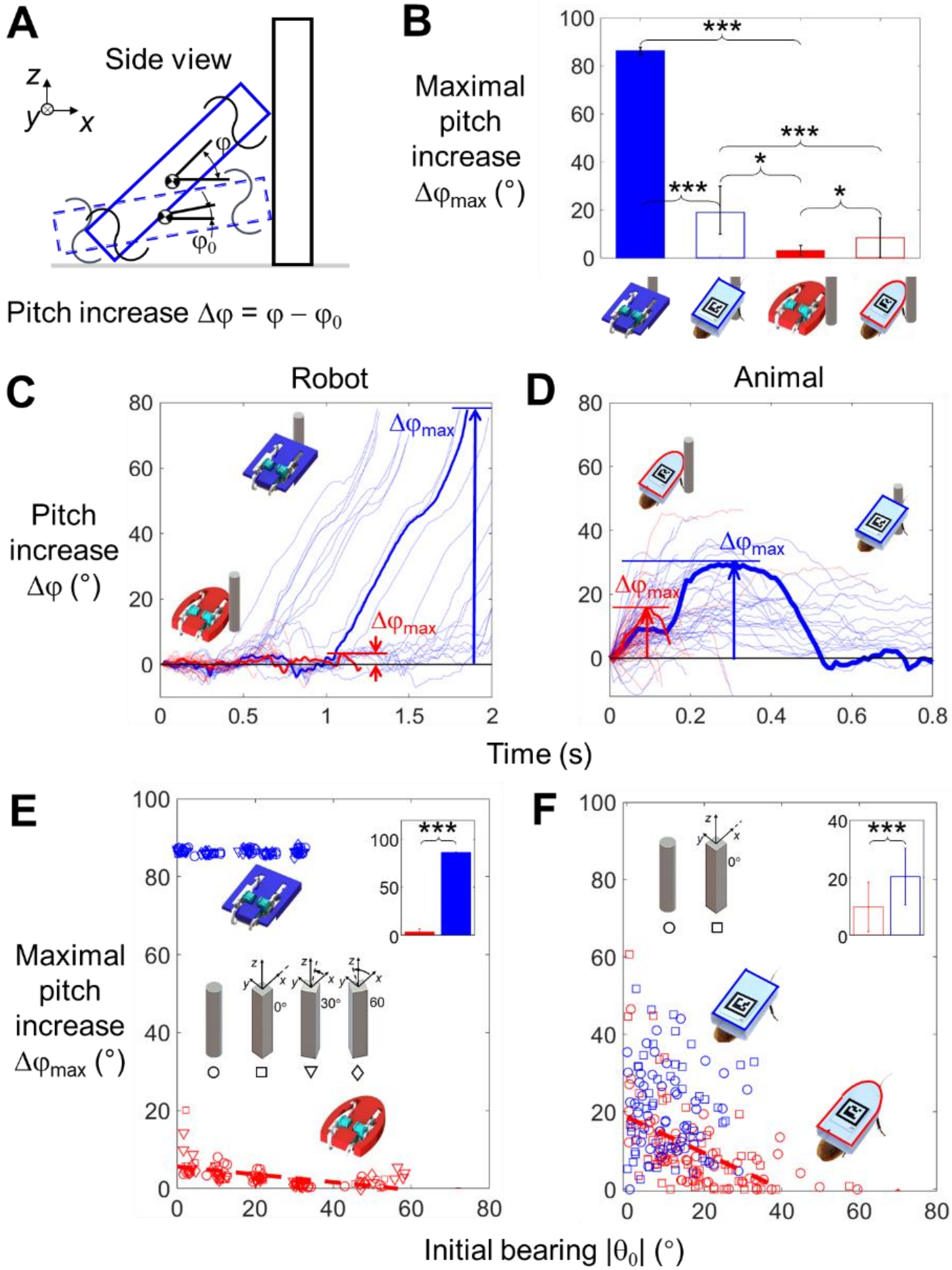

**Fig. 5. Pitching motion during obstacle interaction.** (**A**) Definition of body pitch, φ, and pitch increase, Δφ, during obstacle interaction. Dashed and solid bodies are at initial contact with pitch $\varphi_0$ and a later time.

(**B**) Maximal pitch increase $\Delta\varphi_{max}$ of robot (filled) and animal (open) during interaction. (**C, D**) $\Delta\varphi$ of robot (C) and animal (D) as a function of time during interaction for all trials. Thick curve is a representative trial for each treatment, with arrows showing $\Delta\varphi_{max}$. In (B-D), data are shown for circular pillars and are similar for other pillar shapes and orientations (Fig. 11). (**E, F**) $\Delta\varphi_{max}$ of robot (E) and animal (F) as a function of initial bearing magnitude $|\theta_0|$. Insets show mean maximal pitch increase for cuboidal (blue) and elliptical (red) body shapes. Other definitions follow Fig. 4. In (B, E, F), * and *** show statistical significance with $P < 0.05$ and $P < 0.001$ (Kruskal-Wallis test). See Movie S1 and Movie S2 for representative trials.

Body shape strongly affected body pitching during interaction for both the robot and animal (Fig. 5). With a cuboidal shape, as the body turned towards the obstacle (Fig. 4B) and legs continued to propel, the body often pitched up (Fig. 3A, C, blue curve; Fig. 5C, D, blue; Movie S1). The robot always pitched up (Fig. 3A, blue) to a nearly vertical body orientation (Fig. 5B, solid blue bar) and eventually flipped over (Fig. 3A, black). The animal, however, after initially pitching up (Fig. 3C, blue), always managed to turn sideways to escape (Fig. 3C, yellow) before it could pitch up further and thus never flipped over. As a result, the robot's maximal pitch increase $\Delta\varphi_{max}$ (defined by arrows in Fig. 5C, D) was higher than that of the animal (Fig. 5B, blue; $P < 0.001$, Kruskal-Wallis test). By contrast, with an elliptical shape, as the body was repelled away from the obstacle, $\Delta\varphi_{max}$ was significantly smaller (Fig. 5B, red; Fig. 5E, F; Movie S1) than that with a cuboidal body ($P < 0.001$, Kruskal-Wallis test).

For a cuboidal body, after turning towards the obstacle, large pitching occurred stochastically—the time required for the body to pitch up significantly after initial contact was highly variable from trial to trial (Fig. 5C, D). In some trials, this happened as soon as the body first aligned with the obstacle; in others, this happened after body bearing oscillated several times (Fig. 4C, D).

For both the robot and animal with a cuboidal body, we did not find a significant dependence of $\Delta\varphi_{max}$ on $|\theta_0|$ (Fig. 5E, F, blue; $P > 0.05$, Kruskal-Wallis test). By contrast, for both the robot and animal with an elliptical body, $\Delta\varphi_{max}$ decreased with $|\theta_0|$ during interaction (Fig. 5E, F, red; $P < 0.001$, Kruskal-Wallis test).

### 3.4. Turning and pitching are insensitive to obstacle geometry

Surprisingly, for both the robot and animal, turning and pitching motions were similar for all obstacle shapes and orientations tested (Fig. S5, Fig. S6), with no significant difference in final bearing and pitch increase ($P > 0.05$, Kruskal-Wallis test; Fig. 4E, F, Fig. 5E, F). Final bearing of elliptical robot (Fig. 4E, red) showed a weak dependence on pillar shape ($P = 0.004$, Kruskal-Wallis test) and orientation ($P = 0.02$, Kruskal-Wallis test) but only varied within 2% (mean ± s.d. = 94° ± 2°). This insensitivity is particularly remarkable for the open-loop robot (Movie S2), whose data for different pillar shapes and orientations collapsed onto a straight line (Fig. 4E, Fig. 5E).

## 4. MODELING METHODS

### 4.1. Rationale for potential energy landscape model

For both the robot and animal, body-obstacle interaction consisted of continual collisions due to cyclic self-propulsion (Supplementary Information, Section S4). Although such complex, noisy dynamics can in principle be solved (e.g., using multi-body dynamics simulation (Xuan and Li, 2020b)), such modeling is more like doing an in-silico experiment and may become intractable. Inspired by similarities of our system to robotic grasping, here we chose to use a quasi-static potential energy landscape model to explain how body shape induces obstacle attraction and repulsion during dynamic locomotion, without solving for the dynamics.

Our system is similar to many problems in robotic manipulation (e.g., part alignment (Boothroyd and Ho, 1976; Brost, 1992; Jayaraman, 1996; Peshkin and Sanderson, 1988; Várkonyi, 2014), grasping without force closure (Lynch et al., 1998; Lynch and Mason, 1999; Zumel, 1997)) in that shape strongly affects physical interaction. In addition, in robotic part alignment, motion also consists of continual collisions (Jayaraman, 1996; Mohri and Saito, 1994; Peshkin and Sanderson, 1988; Zumel, 1997). Despite this, quasi-static potential energy field models well explained how part and feeder shapes affect their interactions and informed planning strategies to achieve the desired alignment even with uncertainties in part orientation, friction, and intermittent contact dynamics (Brost, 1992; Peshkin and Sanderson, 1988; Zumel, 1997).

### 4.2. Model approximations and definition

Our potential energy landscape directly results from physical interaction (similar to (Mason et al., 2012)) and is unlike artificial potential fields (Khatib, 1986) or navigation functions (Rimon and Koditschek, 1992). Our model had several simplifying assumptions and approximations. First, we approximated both the locomotor body and obstacle as a rigid body and neglected legs, which rarely contacted the obstacle during interaction. Second, we chose to focus on pitch and bearing (turning) and constrained body roll to be zero considering that it was small during interaction (Fig. S2). In addition, we assumed that the body's lowest point always contacted the ground. Furthermore, the potential energy landscape only modeled the effect of conservative gravitational force. Moreover, we neglected dynamics of the system; thus, prediction of robot's trajectories in dynamic motion is not possible.

The potential energy of the system was $E = mgz$, where $m$ was body mass, $g$ was gravitational acceleration, and $z$ was the body's center of mass height. For each body horizontal position ($x$, $y$) relative to the obstacle, center of mass height and thus potential energy were determined by body pitch and bearing. To obtain the potential energy landscape, we first determined center of mass position and orientation from the reconstructed 3-D motion. For each center of mass position ($x$, $y$, $z$) along the measured trajectory of a trial, we calculated body potential energy as a function of body pitch and bearing by rotating the body about a pitch and a yaw axis through the center of mass. We normalized potential energy to that when body pitch was zero, $mgz_0$, where $z_0$ is center of mass height when body is horizontal (pitch and roll are both zero).

### 4.3. Model analysis of intermediate shapes

To better understand what aspect of body shape induces obstacle attraction and further examine whether there was a critical intermediate body shape between elliptical and cuboidal shape where body-obstacle interaction changed from attractive to repulsive, we calculated the potential energy landscape of superellipses, given by:

$$\left|\frac{x}{a}\right|^n + \left|\frac{y}{b}\right|^n = 1$$

where $2a$ and $2b$ are body length and width, respectively, and $n$ parametrizes the shape ($n = 2$ for ellipse and $n = \infty$ for rectangle). We varied $n$ in the range [2, 20] in increments of 0.25 to create superellipses and extruded them by the body height $2c$ (Fig. 9A). For simplicity, we assumed constant zero bearing. We then moved the body towards the pillar and calculated the potential energy landscape (Fig. 9B, C, D, Movie S7), as described in Section 4.2. We assumed the same mass and uniform mass distribution for all body shapes.

## 5. MODELING RESULTS

### 5.1. Potential energy landscape topology differs between body shapes

Before the body contacted the obstacle, the potential energy landscape was the same for the cuboidal and elliptical bodies (Fig. 6A, D). Potential energy only increased with body pitch and did not change with bearing. When the body was sufficiently close to the obstacle, it must pitch up and/or turn; otherwise, it would penetrate the rigid obstacle, resulting in infinite potential energy (Fig. 6B, E, white). The boundary of the infinite potential energy regions in the landscape represented all possible pitch and bearing states with contact. Hereafter, we refer to these as prohibited regions. As the body moved even closer, it had to turn and/or pitch up more, and the prohibited regions became larger (Fig. 6C, F, white).

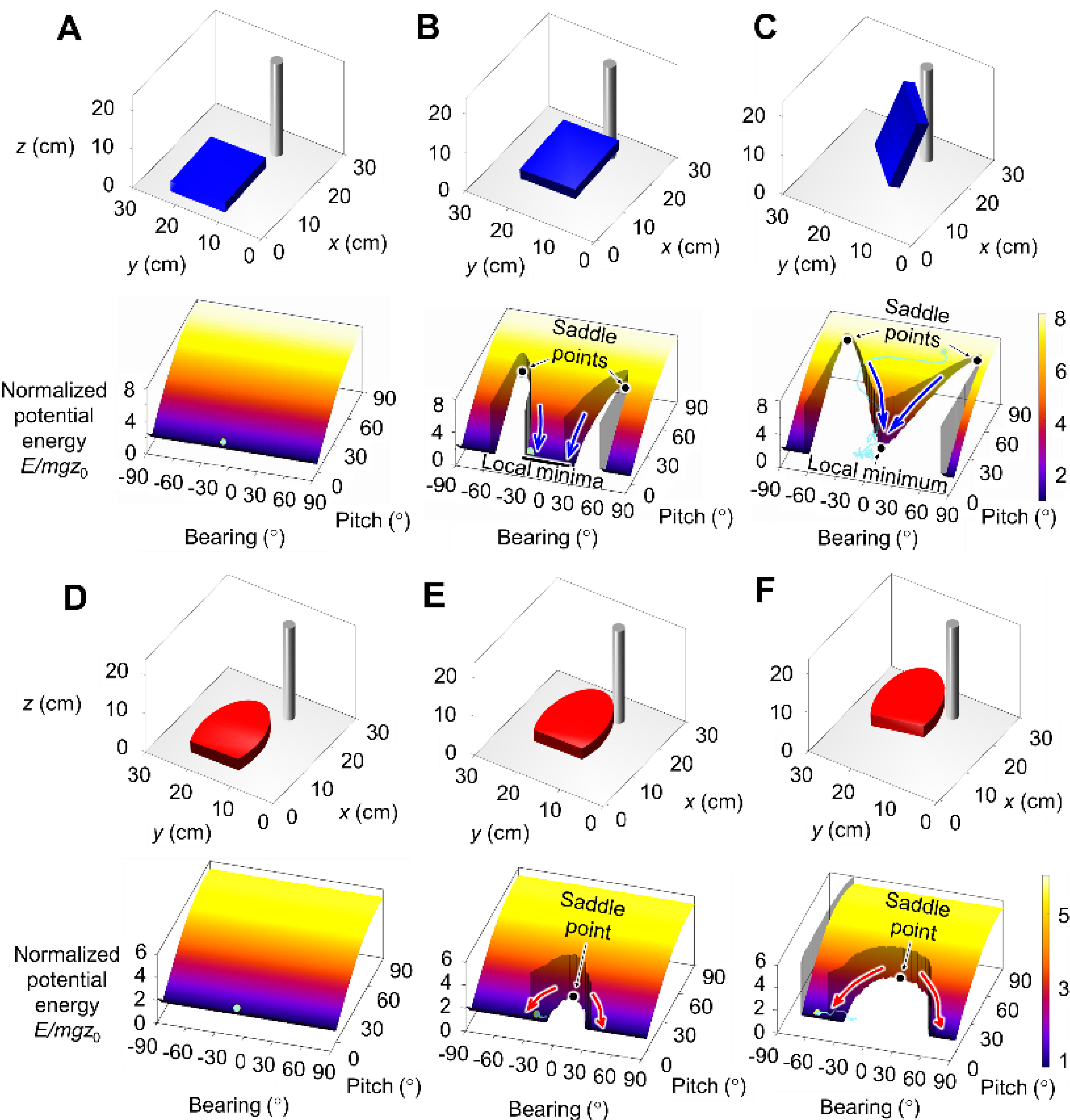

**Fig. 6. Potential energy landscape over body pitch and bearing.** Results are shown for robot body interacting with circular pillars and are similar for animal (Movie S3, Movie S4), other pillar shapes and orientations (Fig. S7), and pillars of large sizes (Fig. 16). (**A**, **B**, **C**) A forward-moving cuboidal body interacting with obstacle results in a landscape that attracts it towards obstacle (bearing converges to zero), which then leads to pitching up. (**D**, **E**, **F**) A forward-moving elliptical body interacting with obstacle results in a landscape that repels it away from obstacle (bearing diverges from zero), with little pitching up. In (A-F), top panel shows body position and orientation; bottom panel shows potential energy landscape. Landscape for negative pitch is symmetric to that of positive pitch and not shown for simplicity. Three snapshots of a representative trial are shown for each shape: (A, D) before contact, (B, E) shortly after contact, and (C, F) as body moves even closer to obstacle. See Movie S3 and Movie S4 for landscape evolution during interaction. White region in landscape encloses impossible states in which body must penetrate rigid obstacle, resulting in prohibited region(s); cyan dot and curve show current system state and state trajectory projected on landscape. Saddle point(s) and local minimum/minima are shown. Blue or red arrows show direction of potential energy gradient along boundary of prohibited region(s).

For the cuboidal body (Movie S3), as it moved close to the obstacle, two prohibited regions emerged (Fig. 6B, C, white), with two saddle points corresponding to frontal bottom corners of the body contacting the obstacle. The inner boundary of prohibited regions between the saddle points corresponded to when the body contacted the obstacle in the front. The outer boundaries beyond the two saddle points corresponded to when the body contacted the obstacle on the sides. When the body contacted the obstacle in the front, body pitch increased with bearing magnitude, i.e., the body must pitch up to turn away from the obstacle. For the elliptical body close to the obstacle (Movie S4), only one prohibited region emerged (Fig. 6E, F, white), centered around zero bearing. The curved boundary of prohibited region corresponded to when the body contacted the obstacle, with the single saddle point being in contact in the front middle. With contact, body pitch decreased with bearing magnitude, i.e., the body pitched down as it turned away from the obstacle.

The different shapes of prohibited regions between the two body shapes resulted in different topologies of attractive basins of the landscape (Fig. 6, B, C vs. E, F, Movie S3 vs. Movie S4). For the cuboidal body, there existed a local minimum basin at zero body pitch around zero bearing between the prohibited regions and two basins at zero body pitch at large body bearing magnitudes beyond the prohibited regions. For the elliptical body, only the latter existed. Hereafter, we refer to the former as the turn-towards-and-pitch basin and the latter as the turn-away basins.

### 5.2. Model explains dependence of turning and pitching on body shape

The distinct topology of attractive landscape basins between the cuboidal and elliptical shapes provided insight into why they resulted in the observed motions. A static body contacting an obstacle can be stable in any orientation as long as friction against the ground and obstacle is sufficient. However, for the self-propelled robot or animal, continual collisions during obstacle interaction (Fig. 4) broke continuous frictional contact and caused the body to be statically unstable. As a result, the system drifted from less (higher energy) to more stable (lower energy) states on the potential energy landscape, i.e., it was attracted to landscape basins.

As the cuboidal body moved closer to the obstacle, the turn-towards-and-pitch basin remained around zero bearing but shrank in the bearing direction and moved up in the positive pitch direction (Fig. 6B, C, Movie S3). As a result of this attractive basin, when the cuboidal body continued to self-propel against the obstacle in the front, it turned towards the obstacle and then pitched up. By contrast, as the elliptical body moved closer, the two pitch-away basins moved away from zero bearing but did not in the pitch direction (Fig. 6E, F, Movie S4). As a result of these two attractive basins, when the elliptical body continued to self-propel against the obstacle in the front, it was repelled away while staying around zero pitch. These findings in self-propelled, dynamic locomotor-obstacle interaction are similar to how feeders are shaped appropriately to trap or reject parts (Brost, 1992; Caine, 1993; Jayaraman, 1996; Peshkin and Sanderson, 1988) and how part shape probabilistically affects its resting orientation when dropped on a feeder (Boothroyd and Ho, 1976; Varkonyi, 2016).

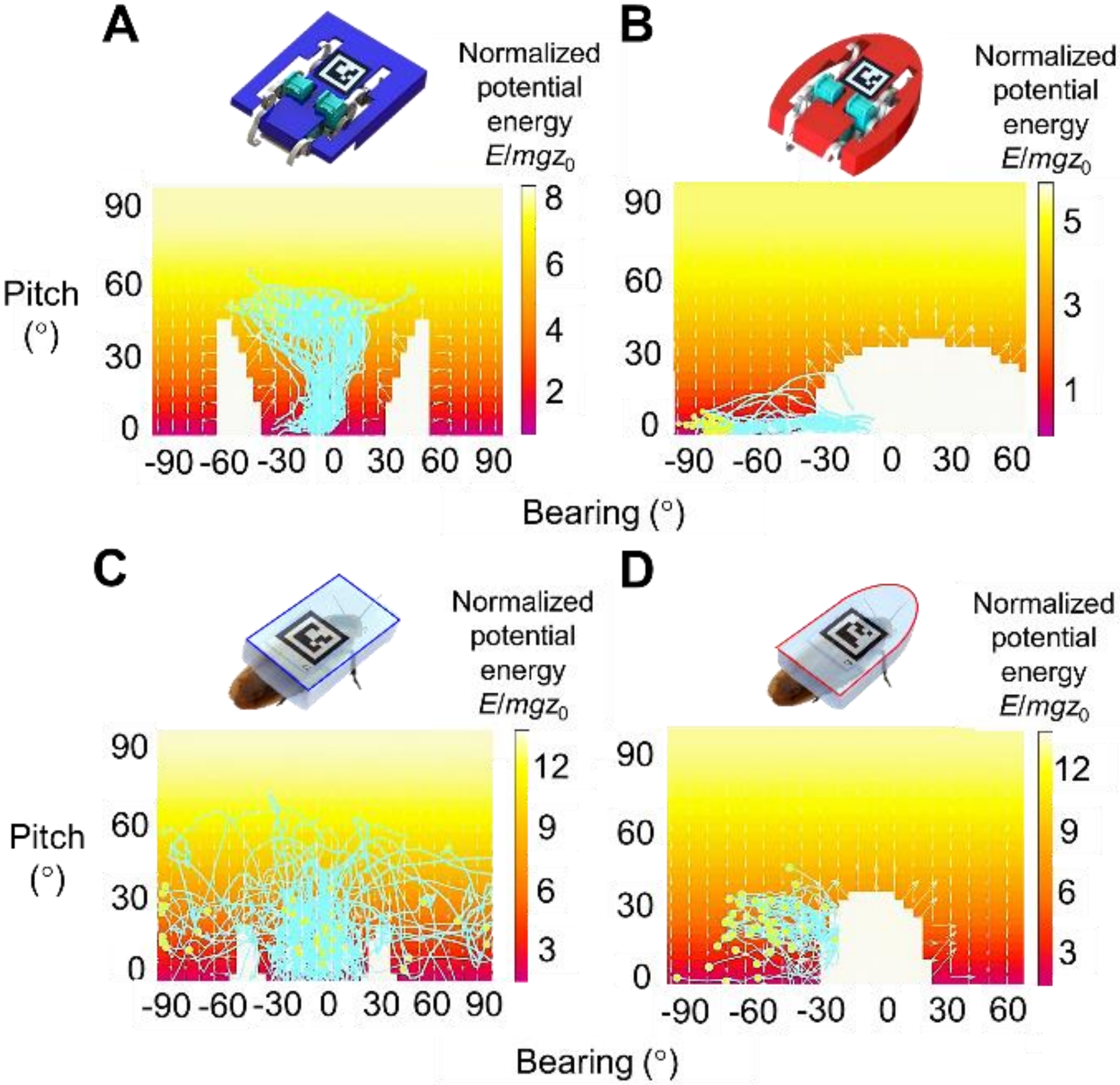

**Fig. 7. Ensemble of system state trajectories on potential energy landscape.** Results are shown for sample trials interacting with pillars with different shapes and orientations. (**A, B**) Robot with cuboidal (A) and elliptical (B) bodies. (**C, D**) Animal with cuboidal (C) and elliptical (D) bodies. For each case, a snapshot of landscape from a representative trial is shown with trajectories of 50 trials (yellow markers and cyan curves), whose initial bearing were closest to the representative trial. See Movie S5 and Movie S6 for landscape and state evolution during interaction. Note that this visualization is an approximation because landscape further depends on body position and thus differs between trials. See Fig. 6 for definition of plots. Trajectories cutting through prohibited region are an artifact from the prohibited region expanding as body moves closer to obstacle as well as using a representative landscape for all trials which have slightly

different landscapes. For D, trials in which the animal turns right are flipped to turning left considering lateral symmetry.

Indeed, the ensemble of system state trajectories from experiments (Fig. 7, Movie S5, Movie S6) showed that, despite the drastic simplification, our potential energy landscape model well explained the dynamic motion of the system resulting from complex, noisy interaction (Fig. S1). With a cuboidal body, the system state in all trials was attracted towards the local minimum basin around zero bearing and then increased in pitch (Fig. 7A, C). With an elliptical body, the system state in all trials was repelled away from zero bearing by the expanding prohibited region (Fig. 7B, D). One exception is that the passive model did not explain how the cuboidal animal escaped obstacle attraction, with system state going out of the attractive basin between the saddle points (Fig. 7C). These quantitative measurements of state trajectories on the landscape also demonstrated that the self-propelled, dynamic system was strongly attracted to the basins but did not reach and stay at the lowest potential energy state.

We note that the observed body turning may be intuitively explained by assessing the torque from body-pillar contact force (which should be approximately perpendicular to the local tangent of the body) in a simple planar model. However, we view potential energy landscape modeling as a general approach to locomotion in complex 3-D terrain (Othayoth et al., 2020). For example, we can use it to study and understand how changes in body or terrain properties (e.g., body shape here, obstacle stiffness in (Othayoth et al., 2020)) affect motion in 3-D, by observing and analyzing how the system state behaved on a landscape over the relevant configuration space of that particular problem (e.g. body bearing and pitch here, body roll and pitch in (Othayoth et al., 2020)).

### 5.3. Model explains insensitivity to obstacle geometry

For both body shapes, the potential energy landscapes from different obstacle shapes and orientations were strikingly similar (Fig. 8, Fig. S7). This explained why they did not affect traversal probability (Fig. 3, Fig. S4) and body turning (Fig. 4, Fig. S5) and pitching (Fig. 5, Fig. S6) motions. In addition, the potential energy landscapes from obstacles with a diameter comparable to or much larger than body size were also qualitatively similar (Fig. 8). Together, these results demonstrated that obstacle attraction followed by pitching up, or obstacle repulsion with little pitching, is an inherent property of the self-propelled locomotor induced by body shape, regardless of obstacle geometry and size.

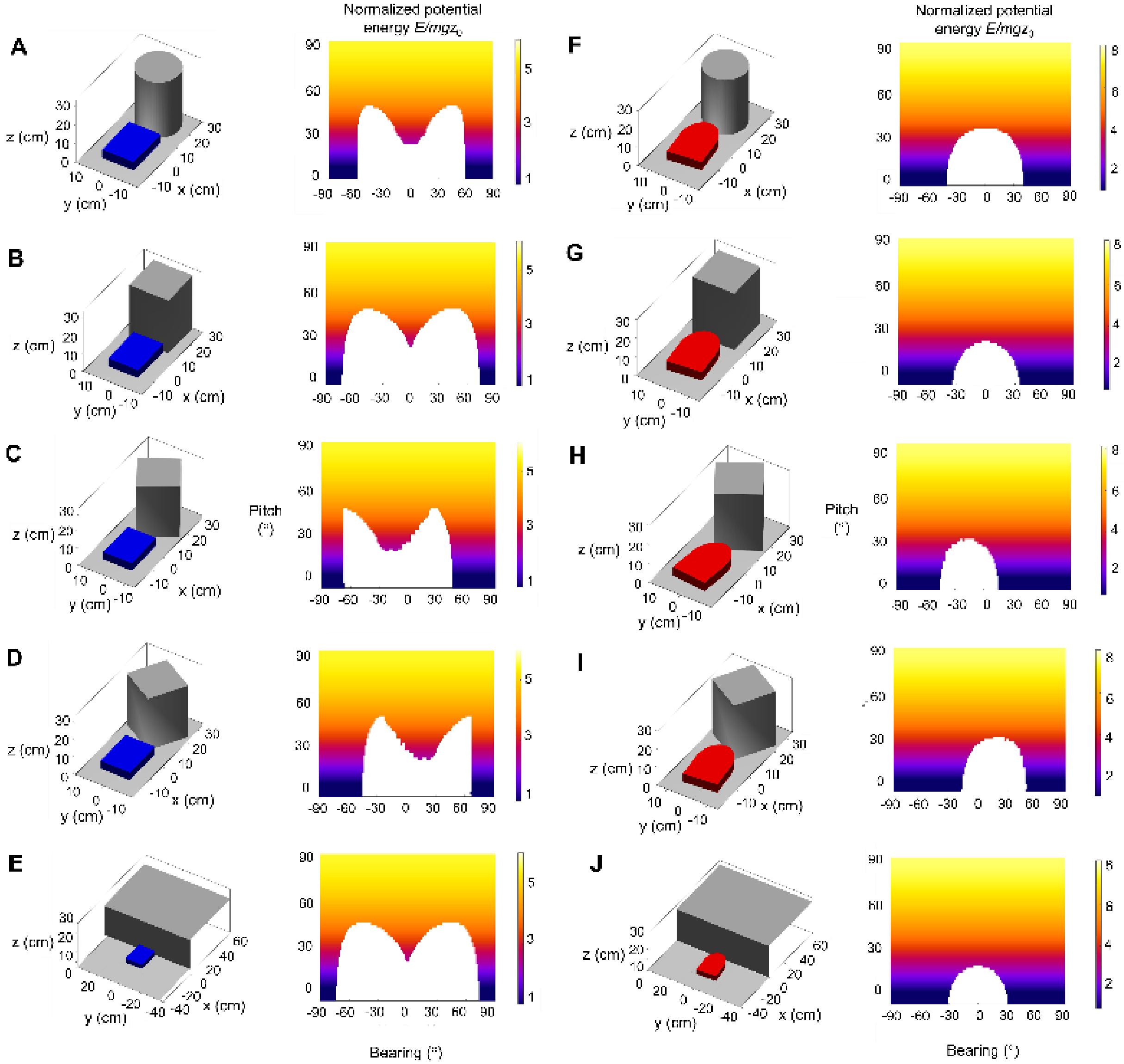

**Fig. 8. Potential energy landscape from large obstacles.** (**A-E**) Cuboidal and **(F-J)** elliptical robot body interacting with a large circular (A, F) and square pillars oriented at 0° (B, G), 30° (C, H), and 60° (D, I) relative to *x*-axis, and with an infinitely large obstacle (E, J). See Fig. 6 for definition of plots.

### 5.4. Model reveals obstacle attraction stems from frontal flatness of body

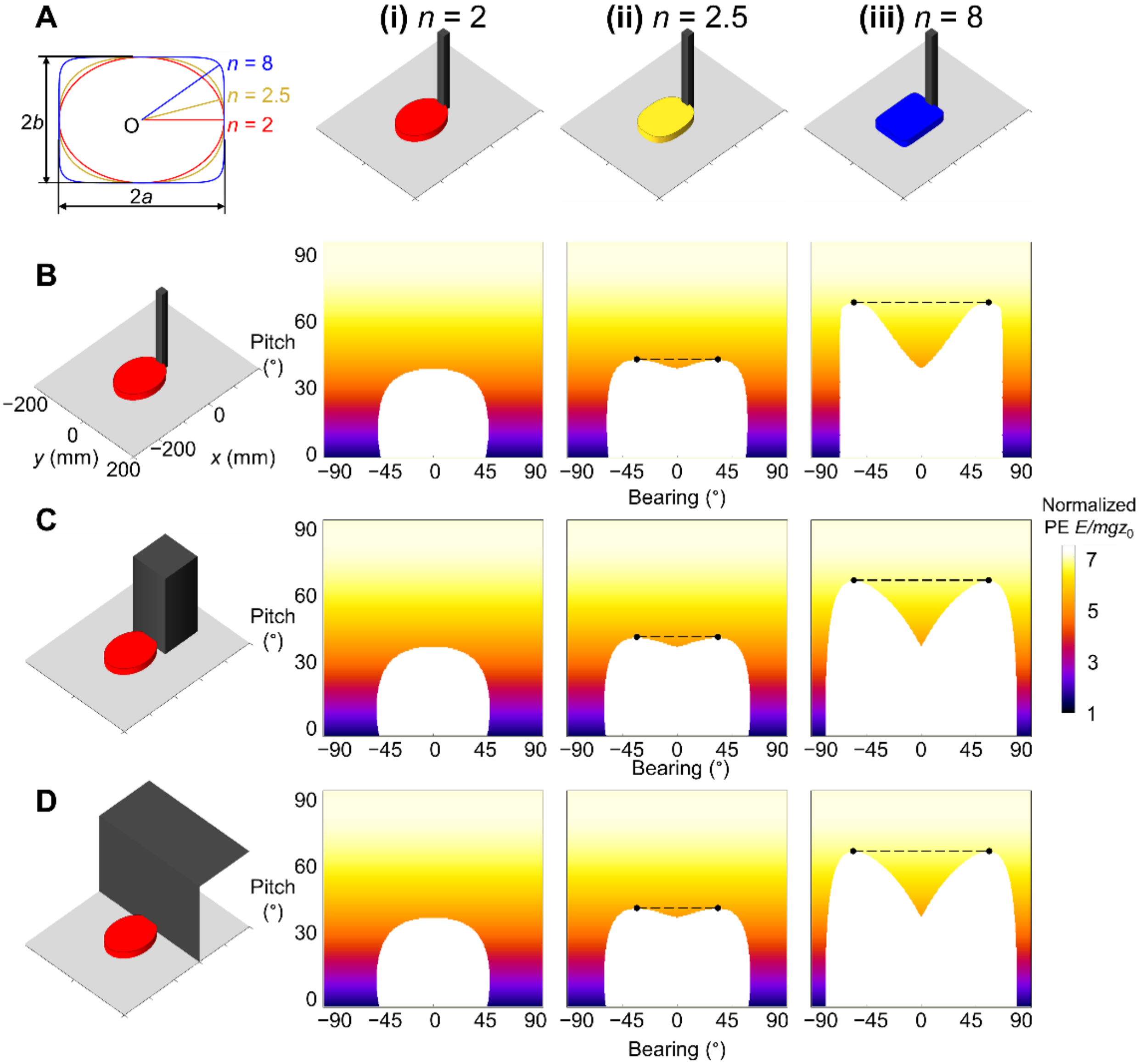

**Fig. 9. Size of attractive basin increases with flatness of front face.** (**A**) Example of superellipse parametrization showing an elliptical shape ($n$ = 2), an intermediate shape ($n$ = 2.5), and a near cuboidal shape ($n$ = 8). Lines from center (O) show maximal radial length of each shape. Maximal radial length of elliptical shape (red line) is the minimal radial length of the other two shapes. (**B, C, D**) Potential energy landscape for a forward-moving body of example shapes (Columns i-iii) with constant zero bearing interacting with pillars of different sizes (left panels, which shows ellipse body as an example). Black dots

in ii and iii show two saddle points and dashed lines and boundaries of prohibited region (white) between saddle points together show the size of attractive basin.

Our model analysis of intermediate shapes showed that obstacle attraction is mainly a result of the flatness of the front of the body and increases as the body becomes more cuboidal. As the body shape changed from elliptical ($n = 2$) to cuboidal ($n > 2$, Fig. 8A, i-iii, Movie S7), the initially curved front face became increasingly flat in the middle. This led to a change in the number of attractive landscape basins, from not having the turn-towards-and-pitch basin to having it, which was insensitive to obstacle size (Fig. 8B, C, D, Movie S7). Because the emergence of the turn-towards-and-pitch basin stems from the flatness of the front face, simply rounding the corners (like in Spot from Boston Dynamics) will not mitigate pitching up and flipping over for cuboidal robots.

We also found that the body shape at which the turn-towards-and-pitch basin emerged depended on body-obstacle distance. In other words, there is not a single "critical" shape that separates the repulsive and attractive regimes. As the body moved very close to the obstacle, all shapes tested became attractive (Movie S7C). This appeared surprising for the elliptical body but was consistent with our experimental observations—the robot and animal were repelled to turn away from the obstacle before they could move so close to it.

## 6. APPLICATIONS

### 6.1. Elliptical body enables open-loop traversal of cluttered obstacles

Our discovery of obstacle repulsion of an elliptical body is useful for passive control of dynamic traversal of cluttered obstacles without sensory feedback. To demonstrate this, we challenged our open-loop robot to traverse a multi-pillar field with narrow gaps (150% body width, pillars arranged in an equilateral triangle grid, Fig. 10A, B, Movie S7), starting from different initial position and body yaw. With a cuboidal body (Fig. 10A), the robot was continually attracted towards every obstacle that it contacted along its way. Thus, it was almost always trapped inside the obstacle field, rarely traversed (Fig. 10C, D, blue), and often flipped over (53% probability within 8 seconds). By contrast, with an elliptical body (Fig. 10B), the robot always traversed the field as it was continually repelled away by obstacles (Fig. 10C, D, red). We also made similar observations in natural terrain cluttered with irregularly shaped boulders (Fig. 10E, F, Movie S8, left; Movie S9, left). Although we only demonstrated this in a small obstacle field, it is almost certain that an elliptical shape can lead to traversal in much larger obstacle fields and could be useful even in aerial or aquatic locomotion (see Supplementary Information, Section S4 and Movie S13).

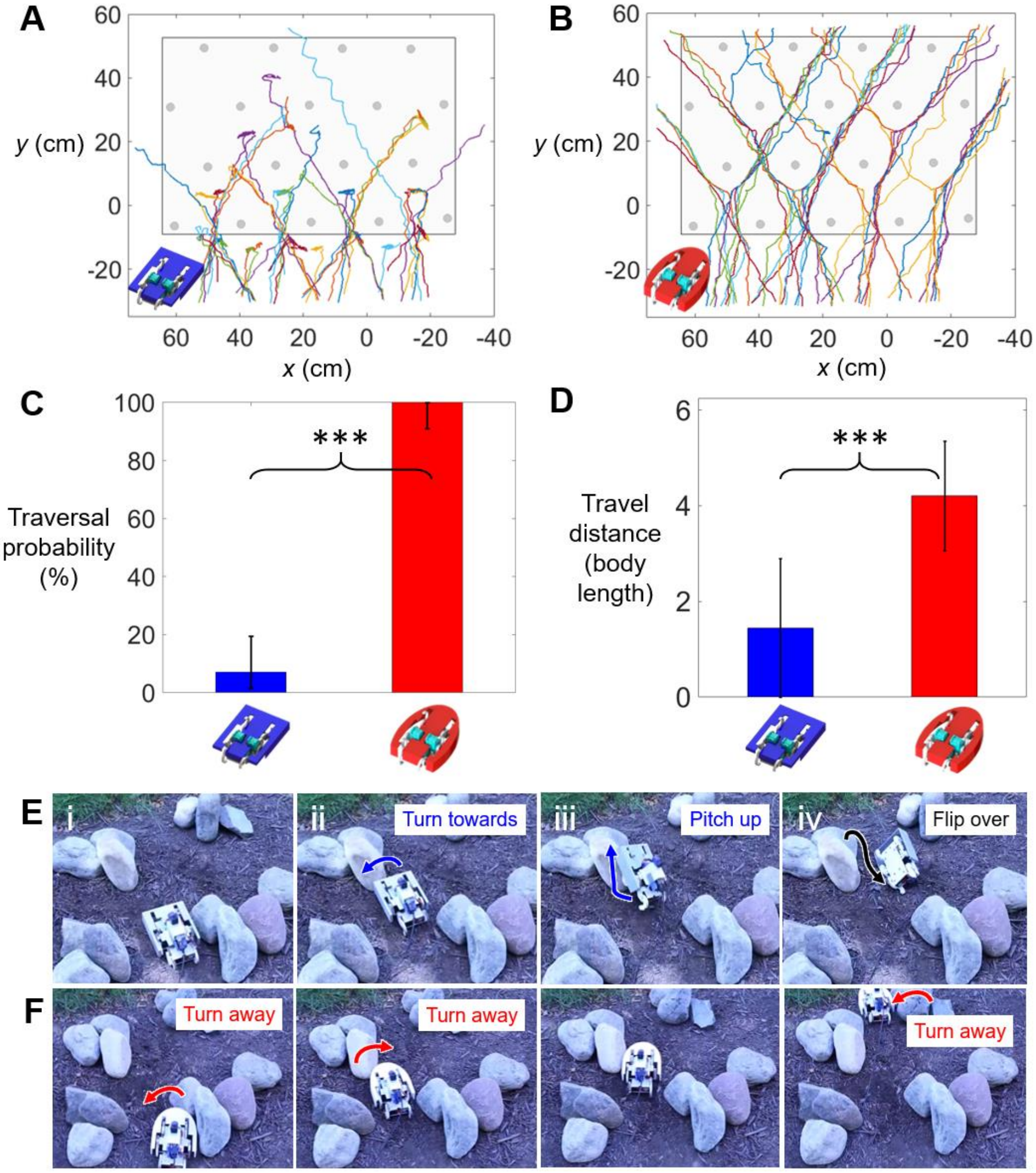

**Fig. 10. Robot traversing cluttered obstacles.** (**A, B**) Robot trajectories in the horizontal plane with a cuboidal (A) and an elliptical (B) body traversing a multi-pillar field, with obstacle spacing of 150% body

length. (**C, D**) Probability of traversing entire field (C) and distance travelled along trajectory within boundary of pillar field (inner gray box in (A, B)) (D) with a cuboidal (blue) and an elliptical (red) body. *** shows statistical significance ($P < 0.001$, Kruskal-Wallis test). (**E, F**) Snapshots of robot with a cuboidal (E) and an elliptical (F) body traversing natural cluttered terrain. Colored arrows show motions defined in Fig. 3. See Movie S7, Movie S8 left, Movie S9 left.

### 6.2. Pitch-and-turn helps cuboidal robots escape obstacle attraction

To inform how cuboidal robots should escape obstacle attraction and avoid subsequent flipping-over, we observed how the animal achieved this. After being attracted towards the obstacle (Fig. 11A, i, ii), the animal often first pushed both hind legs backwards to pitch its body up substantially against the obstacle (Fig. 110A, iii), then extended one hind leg while protracting the other (Fig. 11A, iv), presumably to generate a torque to turn its body away from the obstacle.

Similar to parts overcoming potential energy barriers to reach and remain in a potential energy local minimum state (desired orientation) (Boothroyd and Ho, 1976; Jayaraman, 1996; Varkonyi, 2016; Zumel, 1997), we can view the escape from obstacle attraction as a barrier-crossing transition on our potential energy landscape (Othayoth et al., 2020), escaping the turn-towards-and-pitch basin to reach the turn-away basins. To further understand why the animal pitched up before turning, we examined the potential energy landscape of a forward moving cuboidal body pitching up against the pillar with a constant zero bearing (Movie S11). As the body moved closer and pitched up (Fig. 11B vs. Fig. 11C), potential energy gradient $dE/d\theta$ at the system state along the bearing direction decreased (Fig. 11D). For a body in quasi-static equilibrium, this gradient was the torque exerted by body weight to turn the body towards the pillar. For the self-propelled animal (or robot), because continual collisions break continuous frictional contact, this torque may be similar in magnitude to the dynamically changing torque required to turn against the obstacle. Therefore, the decreasing gradient with increasing pitch suggested that, by pitching its body up, the animal reduced the turning torque required to initiate turning and escape obstacle attraction.

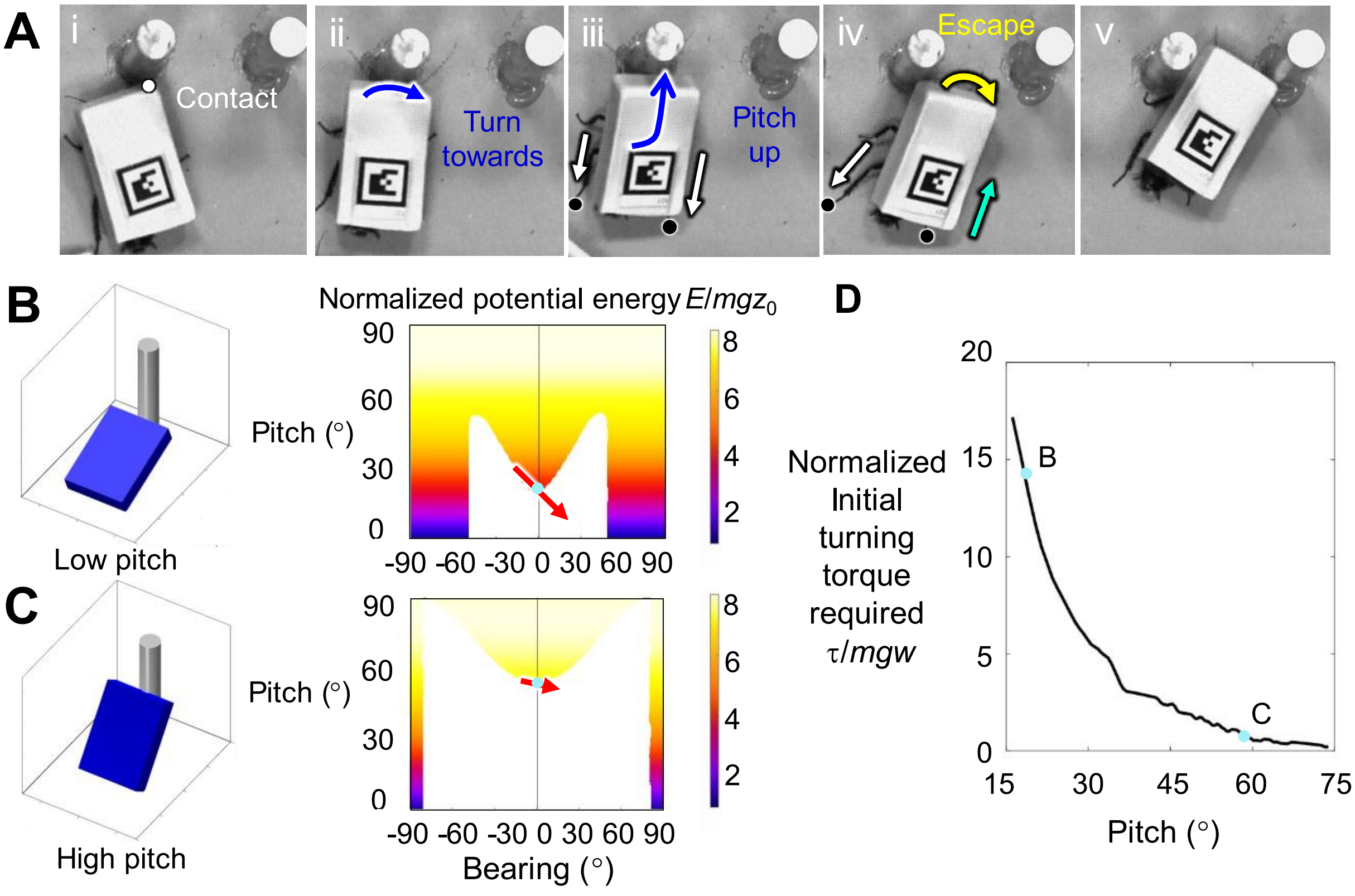

**Fig. 11. Pitching up against an obstacle facilitates escape of a cuboidal body from obstacle attraction.** (**A**) Snapshots of a cuboidal animal escaping after being attracted towards obstacle. Blue and yellow arrows show motions defined in Fig. 3. White and cyan arrows show hind leg extension and protraction. (**B, C**) Potential energy gradient $dE/d\theta$ (red arrow) along bearing direction when body contacts pillar in the front middle (cyan dot) for low (B) and high (C) body pitch, with constant zero bearing. **(D)** $dE/d\theta$ as a function of body pitch, with constant zero bearing. $w$ is half-distance between the hind legs. Cyan markers show values of $dE/d\theta$ from (B, C). See Movie 11 for evolution of landscape and gradient.

Inspired by these animal observations and modeling results, we designed a pitch-and-turn strategy to help cuboidal robots escape obstacle attraction more easily (Movie S12). As the robot approached and pitched up against a pillar obstacle using an alternating tripod gait (Fig. 12A-C, i-iii), we manually triggered the robot to lock its front and mid legs and rotate its hind legs in opposite directions to turn the body until it escaped (Fig. 12A-C, iv) and resumed the alternating tripod gait (Fig. 12A-C, v). In accord with the model prediction of reduced required turning torque with increased pitch (Fig. 11C, D), the robot escaped if its body pitch exceeded a critical value (~ 40º) and otherwise failed to escape due to leg slipping and flipped over. Similarly, the robot traversed natural cluttered terrain using a series of manually triggered pitch-and-turn maneuvers (Fig. 12D; Movie S9, right; Movie S10, right). We emphasize that, although cuboidal robots could simply back up to escape obstacle attraction when obstacles are sparse, active gait change like this is required to traverse highly cluttered terrain (e.g., earthquake rubble).

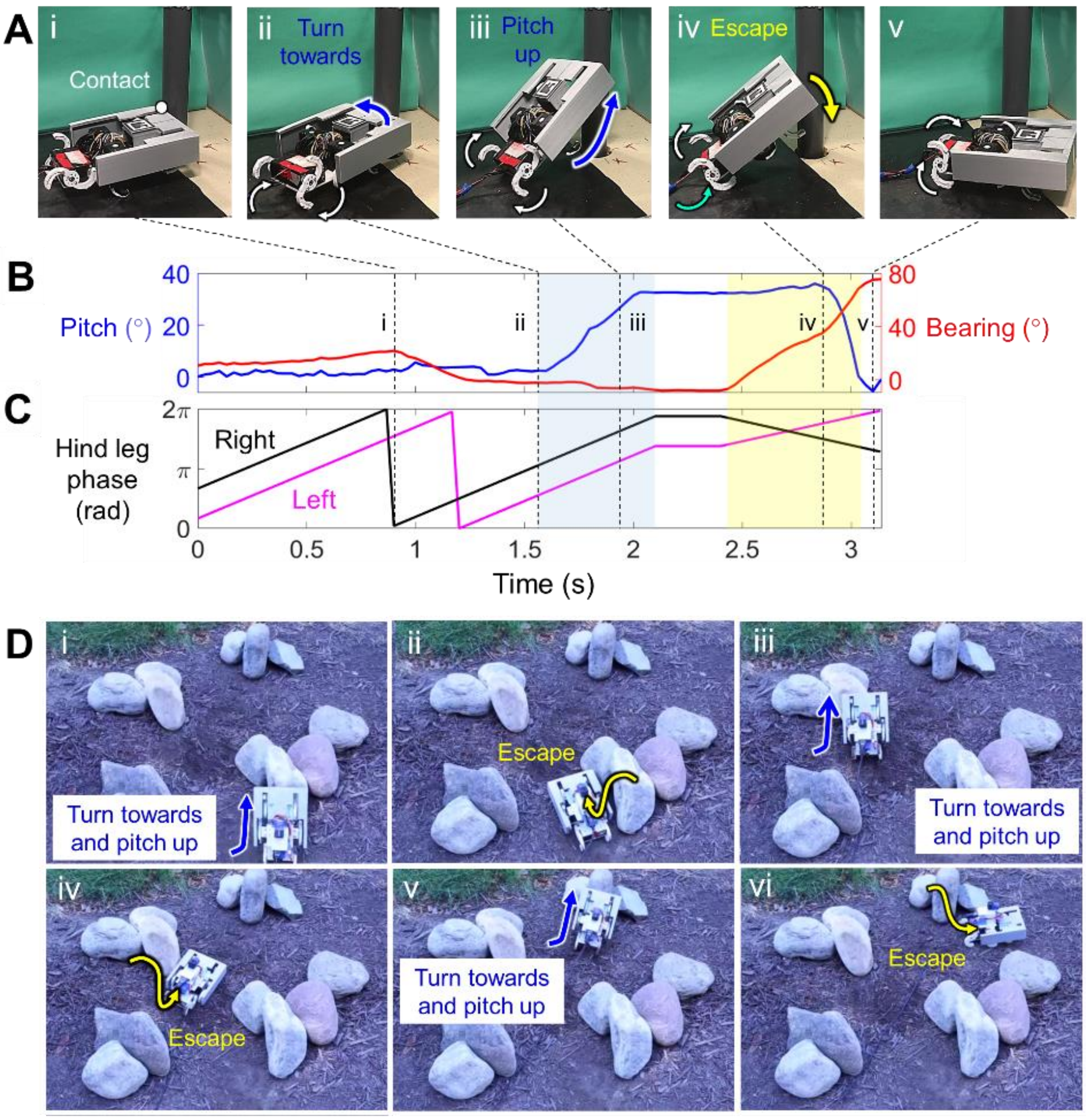

**Fig. 12. Pitch-and-turn strategy helps cuboidal robots escape obstacle attraction.** (**A**) Snapshots of a cuboidal robot using pitch-and-turn strategy to escape attraction by a pillar. Colored arrows correspond to those in Fig. 10A. (**B, C**) Body pitch (blue) and bearing (red) (B) and phases of left (magenta) and right (black) hind legs (C) before, during, and after escape. Gray and green regions show pitching and turning phases. (**D**) A cuboidal robot traversing cluttered boulders using pitch-and-turn strategy. Colored arrows show motions defined in Fig. 3. See Movie S8 right, Movie S9 right.

## 7. DISCUSSION

Our study expands the new concept and usefulness of terradynamic shapes (Li, Pullin, et al., 2015) and is a step in establishing terradynamics (Li et al., 2013) of locomotion in complex 3-D terrain (Gart et al., 2018, 2019; Gart and Li, 2018; Li, Wohrl, et al., 2015; Qian and Koditschek, 2020). We provide new principles of how robots should use body shape to induce obstacle attraction and repulsion to passively control dynamic locomotion. A self-propelled cuboidal body induces attraction towards an obstacle and subsequent pitching up against it, which hinders obstacle traversal but may be useful for initiating climbing. By contrast, a self-propelled elliptical body is repelled away by obstacles and enables rapid traversal without feedback control or planning. This is particularly useful for small robots in time-sensitive or payload-limited missions like search and rescue (Murphy et al., 2008) or extraterrestrial exploration (Bajracharya et al., 2008). More broadly, adopting the right body shape can change the topology of attractive landscape basins and suppress unfavorable locomotor modes. This added to the suite of strategies that robots and animals can use to elicit the desired motions (Gart and Li, 2018; Li et al., 2019; Othayoth et al., 2020, Othayoth et al, in review; Wang et al., 2020; Xuan and Li, 2020a), enabled by a potential energy landscape approach to locomotor transitions (Othayoth et al., 2020).

Our results show that the attractive or repulsive interaction due to locomotor body shape is surprisingly insensitive to the geometry and size of rigid obstacles, the former being similar to those observed in legged robots traversing boulders (Qian and Goldman, 2016). Together, these findings suggest that there may exist general terradynamic principles yet to be uncovered for complex 3-D terrain. We speculate that our findings here also apply to compliant body interacting with compliant obstacles (Othayoth et al., 2020), because changes in stiffness only quantitatively changes the potential energy landscape but the topology is preserved.

Our minimalistic, quasi-static potential energy landscape is surprisingly useful in understanding the probabilistic dependence of locomotion on body shape, which results from complex body-obstacle interaction consisting of many small high-frequency collisions that challenge dynamic modeling. We envision potential energy landscape as the beginning of a statistical physics approach (analogous to (Onuchic and Wolynes, 2004)) to understanding locomotor-terrain interaction probabilistically without solving equations of motion (Othayoth et al., 2020). This is similar to how similar approaches have been successful in robotic grasping. We note that the motions here are only sensitive to the shape of the self-propelled moving body but not sensitive to the geometry of the rigid obstacles, which differs from robotic manipulation where motions are sensitive to the shape of both feeders and parts that move. Considering this, we speculate that physical interactions between two objects are only sensitive to the shape of the body/bodies that move.

We close with a discussion of future directions. First, our study mainly focused on two representative, static body shapes. Discovery of diverse terradynamic shapes, control surfaces, and even morphing shapes (Li et al., 2017; Paik et al., 2012; Steltz et al., 2009) analogous to their aero-/hydrodynamic counterparts (Fish and Lauder, 2017; Lentink et al., 2007; Stowers and Lentink, 2015) to modulate and adjust physical interaction will enable diverse locomotor transitions in robots (Li, Pullin, et al., 2015; Mintchev and Floreano, 2016; Miyashita et al., 2017; Roderick et al., 2017) required for broader applications. This will be facilitated by combining potential energy landscape modeling and composition of complex shapes using parameterizable shape primitives (Rosin, 2000). In addition, studies of interaction with diverse obstacles beyond the rigid pillars tested here will provide terradynamic principles for traversing more heterogeneous terrain. Furthermore, it will be fruitful to model the animal's active behavior (Wang et al., 2020) to understand the principles of how animals, and how robots should, locally sense and react to it (Arslan and Uluç Saranli, 2012) to make the desired locomotor transitions. This will be important in real-world missions with no global knowledge of the environment (Childers et al., 2016). Finally, future dynamic modeling that integrates potential energy landscape (which model conservative forces) with non-conservative forces and stochasticity will enable model-based, probabilistic shape optimization and provide a design tool for robots.

## ACKNOWLEDGMENTS

We thank Dongkai Wang and Tim Greco for help with robot control system development; Yifu Luo and Xiao Yu for assistance during experimental setup and data collection; Tim Greco for help with boulder traversal demonstration; Zheliang Wang for preliminary modeling; Changxin Yan for 3-D reconstruction accuracy verification; Sean Gart, Yucheng Kang, Jean-Michel Mongeau, Mark Cutkosky, Dan Koditschek, Dan Goldman, Bob Full, and Ron Fearing for discussion. This work was funded by a Burroughs Wellcome Fund Career Award at the Scientific Interface, an Arnold & Mabel Beckman Foundation Beckman Young Investigator Award, a U.S. Army Research Office Young Investigator Award under grant number W911NF-17-1-0346, and The Johns Hopkins University Whiting School of Engineering start-up funds to C.L.

## AUTHOR CONTRIBUTIONS

Y.H. designed study, developed robot and animal experimental setup, performed robot experiments, analyzed data, developed model, and wrote an early manuscript; R.O performed model analysis for superellipses and revised paper; Y.W. assisted robot development and experiments and developed pitched turning strategy; C.C.H and R.d.l.T.O. performed animal experiments; E.F. performed drone experiments; C.L. designed and supervised the study, defined data and model analysis, assisted model development, and wrote and revised paper.

## REFERENCES

Ackerman E (2016) Boston Dynamics' SpotMini Is All Electric, Agile, and Has a Capable Face-Arm.

Arslan Ö and Saranli U (2012) Reactive planning and control of planar spring-mass running on rough terrain. *IEEE Transactions on Robotics* 28(3): 567–579. DOI: 10.1109/TRO.2011.2178134.

Arslan Ö and Saranli U (2012) Reactive Planning and Control of Planar Spring–Mass Running on Rough Terrain. *IEEE Transactions on Robotics* 28(3): 567–579. DOI: 10.1109/TRO.2011.2178134.

Baisch AT and Wood RJ (2011) Design and fabrication of the Harvard ambulatory micro-robot. In: *Robotics Research*. Springer, Berlin, Heidelberg, pp. 715–730.

Bajracharya M, Maimone MW and Helmick D (2008) Autonomy for Mars Rovers: Past, present, and future. *Computer* 41(12): 44–50. DOI: 10.1109/MC.2008.479.

Bekker MG (1956) *Theory of Land Locomotion: The Mechanics of Vehicle Mobility*. University of Michigan Press.

Bell WJ, Roth LM and Nalepa CA (2007) *Cockroaches: Ecology, Behavior, and Natural History*. Johns Hopkins University Press.

Birkmeyer P, Peterson K and Fearing RS (2009) DASH: A dynamic 16g hexapedal robot. In: *Intelligent Robots and Systems, 2009. IROS 2009. IEEE/RSJ International Conference on*, 2009, pp. 2683–2689.

Blackman DJ, Nicholson J V., Ordonez C, Miller BD and Clark JE (2016) Gait development on Minitaur, a direct drive quadrupedal robot. *Unmanned Systems Technology XVIII* Karlsen RE, Gage DW,

Shoemaker CM, et al. (eds) 9837: 98370I. DOI: 10.1117/12.2231105.

Boothroyd G and Ho C (1976) Natural Resting Aspects of Parts for Automatic Handling. *American Society of Mechanical Engineers (Paper)* (76-WA/Prod-40): 314–317.

Borenstein J and Koren Y (1991) The Vector Field Histogram - Fast Obstacle Avoidance for Mobile Robots. *IEEE Journal of Robotics and Automation* 7(3): 278–288. DOI: 10.1109/70.88137.

Brost RC (1992) Dynamic analysis of planar manipulation tasks. In: *Proceedings 1992 IEEE International Conference on Robotics and Automation*, 1992, pp. 2247–2254. IEEE Comput. Soc. Press. DOI: 10.1109/ROBOT.1992.219924.

Caine ME (1993) The Design of Shape from Motion Constraints.: 1–247.

Childers M, Lennon C, Bodt B, Pusey J, Hill S, Camden R, Oh J, Dean R, Diberardino C and Karumanchi S (2016) US Army Research Laboratory ( ARL ) Robotics Collaborative Technology Alliance 2014 Capstone Experiment.

Crall JD, Gravish N, Mountcastle AM and Combes SA (2015) BEEtag: A low-cost, image-based tracking system for the study of animal behavior and locomotion. *PLoS ONE* 10(9): 1–13. DOI: 10.1371/journal.pone.0136487.

De A and Koditschek DE (2018) *Vertical Hopper Compositions for Preflexive and Feedback-Stabilized Quadrupedal Bounding, Pacing, Pronking, and Trotting*. DOI: 10.1177/0278364918779874.

Dickinson MH, Farley CT, Full RJ, Koehl MAR, Kram R and Lehman S. (2000) How Animals Move: An Integrative View. *Science* 288(5463): 100–106. DOI: 10.1126/science.288.5463.100.

Dissanayake MWMG, Newman P, Clark S, Durrant-Whyte HF and Csorba M (2001) A solution to the simultaneous localization and map building (SLAM) problem. *IEEE Transactions on Robotics and Automation* 17(3): 229–241. DOI: 10.1109/70.938381.

Elfes A (1989) Using Occupancy Grids for Mobile Robot Perception and Navigation. *Computer* 22(6):

46–57. DOI: 10.1109/2.30720.

Fish FE and Lauder G V (2017) Control surfaces of aquatic vertebrates: active and passive design and function. *The Journal of Experimental Biology* 220(23): 4351–4363. DOI: 10.1242/jeb.149617.

Full RJ and Koditschek DE (1999) Templates and anchors: neuromechanical hypotheses of legged locomotion on land. *The Journal of Experimental Biology* 2(12): 3–125.

Full RJ, Kubow T, Schmitt J, et al. (2006) Quantifying Dynamic Stability and Maneuverability in Legged Locomotion. *Integrative and Comparative Biology* 42(1): 149–157. DOI: 10.1093/icb/42.1.149.

Garcia Bermudez FL, Julian RC, Haldane DW, Abbeel P and Fearing RS (2012) Performance analysis and terrain classification for a legged robot over rough terrain. *IEEE International Conference on Intelligent Robots and Systems*: 513–519. DOI: 10.1109/IROS.2012.6386243.

Gart SW and Li C (2018) Body-terrain interaction affects large bump traversal of insects and legged robots. *Bioinspiration & Biomimetics* 13(2): 026005. DOI: 10.1088/1748-3190/aaa2d0.

Gart SW, Yan C, Othayoth R, Ren Z and Li C (2018) Dynamic traversal of large gaps by insects and legged robots reveals a template. *Bioinspiration and Biomimetics* 13(2). IOP Publishing: aaa2cd. DOI: 10.1088/1748-3190/aaa2cd.

Gart SW, Mitchel TW and Li C (2019) Snakes partition their body to traverse large steps stably. *The Journal of Experimental Biology* 222(8): jeb185991. DOI: 10.1242/jeb.185991.

Guizzo E and Ackerman E (2015) The Hard Lessons of DARPA ' s Robotics Challenge. *IEEE Spectrum* 52(8): 11–13.

Haldane DW and Fearing RS (2015) Running beyond the bio-inspired regime. In: *Proceedings - IEEE International Conference on Robotics and Automation*, 2015, pp. 4539–4546. DOI: 10.1109/ICRA.2015.7139828.

Hedrick TL (2008) Software techniques for two- and three-dimensional kinematic measurements of biological and biomimetic systems. *Bioinspiration & Biomimetics* 3(3): 034001. DOI: 10.1088/1748-3182/3/3/034001.

Hutter M, Gehring C, Bloesch M, Hoepflinger MA, Remy CD and Siegwart R (2012) Starleth: a compliant quadrupedal robot for fast, efficient, and versatile locomotion. In: *Adaptive Mobile Robotics*, pp. 483–490. DOI: 10.1142/9789814415958_0062.

Jayaraman K (1996) *Mechanics of entrapment with application to design of industrial parts feeders*. Massachusetts Institute of Technology. Available at: https://dspace.mit.edu/handle/1721.1/10558.

Khatib O (1986) Real-Time Obstacle Avoidance for Manipulators and Mobile Robots. *Autonomous Robot Vehicles*. New York, NY: Springer New York: 396–404. DOI: 10.1007/978-1-4613-8997-2_29.

Kim S, Clark JE and Cutkosky MR (2006) ISprawl: Design and tuning for high-speed autonomous open-loop running. *International Journal of Robotics Research* 25(9): 903–912. DOI: 10.1177/0278364906069150.

Latombe J-C (2012) *Robot Motion Planning*. Springer Science & Business Media.

Lavalle SM (1998) Rapidly-Exploring Random Trees: A New Tool for Path Planning. DOI: 10.1.1.35.1853.

Lentink D, Müller UK, Stamhuis EJ, De Kat R, Van Gestel W, Veldhuis LLM, Henningsson P, Hedenström A, Videler JJ and Van Leeuwen JL (2007) How swifts control their glide performance with morphing wings. *Nature* 446(7139). Nature Publishing Group: 1082–1085. DOI: 10.1038/nature05733.

Leonard JJ and Durrant-Whyte HF (1991) Mobile robot localization by tracking geometric beacons. *IEEE Transactions of Robotics and Automation* 7(3): 376–382.

Li C, Umbanhowar PB, Komsuoglu H, Koditschek DE and Goldman DI (2009) Sensitive dependence of the motion of a legged robot on granular media. 106(9): 3029–3034.

Li C, Zhang T and Goldman DI (2013) A terradynamics of legged locomotion on granular media. *Science* 339(6126): 1408–1412. DOI: 10.1126/science.1229163.

Li C, Wohrl T, Lam H, et al. (2015) Self-righting behavior of cockroaches. *APS March Meeting 2015, abstract #B47.003*. Available at: http://adsabs.harvard.edu/abs/2015APS..MARB47003L (accessed 6 December 2017).

Li C, Pullin AO, Haldane DW, et al. (2015) Terradynamically streamlined shapes in animals and robots enhance traversability through densely cluttered terrain. *Bioinspiration and Biomimetics* 10(4). IOP Publishing: 46003. DOI: 10.1088/1748-3190/10/4/046003.

Li C, Kessens CC, Young A, Fearing RS and Full RJ (2016) Cockroach-inspired winged robot reveals principles of ground-based dynamic self-righting. In: *IEEE International Conference on Intelligent Robots and Systems*, 2016, pp. 2128–2134. DOI: 10.1109/IROS.2016.7759334.

Li C, Kessens CC, Fearing RS and Full RJ (2017) Mechanical principles of dynamic terrestrial self-righting using wings. *Advanced Robotics*. DOI: 10.1080/01691864.2017.1372213.

Li C, Wöhrl T, Lam HK and Full RJ (2019) Cockroaches use diverse strategies to self-right on the ground. *The Journal of Experimental Biology* 222(15): jeb186080. DOI: 10.1242/jeb.186080.

Lynch KM and Mason MT (1999) Dynamic Nonprehensile Manipulation: Controllability, Planning, and Experiments. *The International Journal of Robotics Research* 18(1): 64–92. DOI: 10.1177/027836499901800105.

Lynch KM, Shiroma N, Arai H and Tanie K (1998) The roles of shape and motion in dynamic manipulation: The butterfly example. *Proceedings - IEEE International Conference on Robotics and Automation* 3: 1958–1963. DOI: 10.1109/ROBOT.1998.680600.

Mason MT, Rodriguez A, Srinivasa SS and Vazquez AS (2012) Autonomous manipulation with a general-purpose simple hand. *The International Journal of Robotics Research* 31(5). SAGE PublicationsSage UK: London, England: 688–703. DOI: 10.1177/0278364911429978.

Mintchev S and Floreano D (2016) Adaptive morphology: A design principle for multimodal and multifunctional robots. *IEEE Robotics and Automation Magazine* 23(3). IEEE: 42–54. DOI: 10.1109/MRA.2016.2580593.

Miyashita S, Guitron S, Li S and Rus D (2017) Robotic metamorphosis by origami exoskeletons. *Science Robotics* 2(10): eaao4369. DOI: 10.1126/scirobotics.aao4369.

Mohri N and Saito N (1994) Some effects of ultrasonic vibration on the inserting operation. *The International Journal of Advanced Manufacturing Technology* 9(4): 225–230. DOI: 10.1007/BF01751120.

Murphy RR, Tadokoro S, Nardi D, Jacoff A, Fiorini P, Choset H and Erkmen AM (2008) Search and Rescue Robotics. *Springer handbook of robotics*: 1151–1173.

Onuchic JN and Wolynes PG (2004) Theory of protein folding. *Current Opinion in Structural Biology* 14(1): 70–75. DOI: 10.1016/j.sbi.2004.01.009.

Ordonez C, Shill J, Johnson A, Clark J and Collins E (2013) Terrain identification for RHex-type robots. In: *Unmanned Systems Technology XV* (eds RE Karlsen, DW Gage, CM Shoemaker, et al.), May 2013, p. 87410Q. DOI: 10.1117/12.2016169.

Othayoth R, Thoms G and Li C (2020) An energy landscape approach to locomotor transitions in complex 3D terrain. *Proceedings of the National Academy of Sciences* 117(26): 14987–14995. DOI: 10.1073/pnas.1918297117.

Othayoth R, Xuan Q and Li C (n.d.) Appendage flailing facilitates strenuous ground self-righting. *eLife* in review(in review).

Pacejka HB (2005) *Tyre and Vehicle Dynamics*. Elsevier. DOI: 10.1016/B978-0-08-097016-5.00011-5.

Paik JK, An B, Rus D and Wood RJ (2012) Robotic origamis: self-morphing modular robots. *Proc. 2nd Int. Conf. on Morphological Computation*: 29–31.

Peshkin MA and Sanderson AC (1988) Planning Robotic Manipulation Strategies for Workpieces that Slide. *IEEE Journal on Robotics and Automation* 4(5): 524–531. DOI: 10.1109/56.20437.

Pierre R St. and Bergbreiter S (2016) Gait Exploration of Sub-2 g Robots Using Magnetic Actuation. *IEEE Robotics and Automation Letters* 2(1): 34–40. DOI: 10.1109/lra.2016.2523603.

Qian F and Goldman D (2016) The dynamics of legged locomotion in heterogeneous terrain: universality in scattering and sensitivity to initial conditions. In: *Robotics: Science and Systems*, 2016. DOI: 10.15607/rss.2015.xi.030.

Qian F and Koditschek DE (2020) An obstacle disturbance selection framework: emergent robot steady states under repeated collisions. *International Journal of Robotics Research*. DOI: 10.1177/0278364920935514.

Qian F, Ramesh D and Koditschek D (2019) Towards Obstacle-aided Legged Locomotion in Cluttered Environments. *Bulletin of the American Physical Society*. American Physical Society.

Raibert M (2008) BigDog, the rough-terrain quadruped robot. In: *IFAC Proceedings Volumes (IFAC-PapersOnline)*, 2008. DOI: 10.3182/20080706-5-KR-1001.4278.

Rieser JM, Schiebel PE, Pazouki A, Qian F, Goddard Z, Wiesenfeld K, Zangwill A, Negrut D and Goldman DI (2019) Dynamics of scattering in undulatory active collisions. *Physical Review E* 99(2). American Physical Society: 17–19. DOI: 10.1103/PhysRevE.99.022606.

Rimon E and Koditschek DE (1992) Exact robot navigation using artificial potential functions. *IEEE Transactions on Robotics and Automation* 8(5): 501–518. DOI: 10.1109/70.163777.

Roderick WRTT, Cutkosky MR and Lentink D (2017) Touchdown to take-off: at the interface of flight and surface locomotion. *Interface Focus* 7(1): 20160094. DOI: 10.1098/rsfs.2016.0094.

Rosin PL (2000) Fitting superellipses. *IEEE Transactions on Pattern Analysis and Machine Intelligence* 22(7): 726–732. DOI: 10.1109/34.865190.

Saranli U, Buehler M and Koditschek DE (2001) RHex: A Simple and Highly Mobile Hexapod Robot. *Int. J. of Robotics Research* 20(7): 616–631.

Spagna JC, Goldman DI, Lin P-C, et al. (2007) Distributed mechanical feedback in arthropods and robots simplifies control of rapid running on challenging terrain. *Bioinspiration & Biomimetics* 2(1): 9–18. DOI: 10.1088/1748-3182/2/1/002.

Sponberg S and Full RJ (2008) Neuromechanical response of musculo-skeletal structures in cockroaches during rapid running on rough terrain.: 433–446. DOI: 10.1242/jeb.012385.

Steltz E, Mozeika A, Rodenberg N, Brown E, and Jaeger HM (2009) JSEL: Jamming skin enabled locomotion. In: *2009 IEEE/RSJ International Conference on Intelligent Robots and Systems, IROS 2009*, October 2009, pp. 5672–5677. IEEE. DOI: 10.1109/IROS.2009.5354790.

Stowers AK and Lentink D (2015) Folding in and out: passive morphing in flapping wings. *Bioinspiration & Biomimetics* 10(2). IOP Publishing: 025001. DOI: 10.1088/1748-3190/10/2/025001.

Thrun S, Burgard W and Fox D (2000) A real-time algorithm for mobile robot mapping with applications to multi-robot and 3D mapping. In: *Proceedings 2000 ICRA. Millennium Conference. IEEE International Conference on Robotics and Automation. Symposia Proceedings (Cat. No.00CH37065)*, 2000, pp. 321–328. IEEE. DOI: 10.1109/ROBOT.2000.844077.

Ting LH, Blickhan R and Full RJ (1994) Dynamic and Static stability in hexapedal runners. 269: 251–269.

Transeth AA, Leine RI, Glocker C, Pettersen KY and Liljebäck P (2008) Snake robot obstacle-aided locomotion: Modeling, simulations, and experiments. *IEEE Transactions on Robotics* 24(1): 88–104. DOI: 10.1109/TRO.2007.914849.

Tribelhorn B and Dodds Z (2007) Evaluating the roomba: A low-cost, ubiquitous platform for robotics research and education. *Proceedings - IEEE International Conference on Robotics and Automation*

(April): 1393–1399.

Varkonyi PL (2016) Sensorless Part Feeding with Round Cages in Two and Three Dimensions. *IEEE Robotics and Automation Letters* 1(2): 724–731. DOI: 10.1109/LRA.2016.2521933.

Várkonyi PL (2014) Estimating part pose statistics with application to industrial parts feeding and shape design: New metrics, algorithms, simulation experiments and datasets. *IEEE Transactions on Automation Science and Engineering* 11(3): 658–667. DOI: 10.1109/TASE.2014.2318831.

Wang Y, Othayoth R and Li C (2020) Active adjustments help cockroaches traverse obstacles by lowering potential energy barrier. In: *APS March Meeting 2020, abstract #S22.00003*, Denver, CO, 2020. Available at: https://meetings.aps.org/Meeting/MAR20/Session/S22.3.

Wong JY (2008) *Theory of Ground Vehicles*. John Wiley & Sons.

Xuan Q and Li C (2020a) Coordinated Appendages Accumulate More Energy to Self-Right on the Ground. *IEEE Robotics and Automation Letters* 5(4): 6137–6144. DOI: 10.1109/LRA.2020.3011389.

Xuan Q and Li C (2020b) Randomness in appendage coordination facilitates strenuous ground self-righting. *Bioinspiration & Biomimetics* 21(1): 1–9. DOI: 10.1088/1748-3190/abac47.

Zumel NB (1997) *A Nonprehensile Method for Reliable Parts Orienting*. Carnegie Mellon University.

Supplementary Information for

**Shape-induced obstacle attraction and repulsion during dynamic locomotion**

Yuanfeng Han, Ratan Othayoth, Yulong Wang, Chun-Cheng Hsu, Rafael de la Tijera Obert, Evains Francois, Chen Li*

*Corresponding author (chen.li@jhu.edu)

**This PDF file includes**:

Tables S1 to S3

Supplementary Text S1 to S5

Supplementary References

Supplementary Figures S1 to S8

Legends for Supplementary Movies S1 to S13

**Other supplementary materials for this manuscript include the following:**

Supplementary Movies S1 to S13

## Supplementary Tables

**Table S1. Sample size, mass, and dimensions of robot experiments.** Average data are mean ± s.d.

<table>
<tr><th>Body shape experiments</th><th>Cuboidal</th><th>Elliptical</th></tr>
<tr><td>Total number of trials</td><td>100</td><td>100</td></tr>
<tr><td>Number of trials for circular pillar</td><td>25</td><td>25</td></tr>
<tr><td>Number of trials for 0° square pillar</td><td>25</td><td>25</td></tr>
<tr><td>Number of trials for 30° square pillar</td><td>25</td><td>25</td></tr>
<tr><td>Number of trials for 60° square pillar</td><td>25</td><td>25</td></tr>
<tr><td>Body mass (g)</td><td>219</td><td>226</td></tr>
<tr><td>Body length (cm)</td><td colspan="2">16.8</td></tr>
<tr><td>Body width (cm)</td><td colspan="2">12.8</td></tr>
<tr><td>Body thickness (cm)</td><td colspan="2">2.5</td></tr>
<tr><td>Pillar cross section diameter/side length (cm)</td><td colspan="2">2.5</td></tr>
<tr><td>Pillar height (cm)</td><td colspan="2">30</td></tr>
<tr><th>Multi-pillar experiments</th><th>Cuboidal</th><th>Elliptical</th></tr>
<tr><td>Number of trials</td><td>64</td><td>64</td></tr>
<tr><td>Pillar spacing (cm)</td><td colspan="2">19</td></tr>
<tr><th>Drone experiments</th><th>Cuboidal</th><th>Elliptical</th></tr>
<tr><td>Body width (cm)</td><td colspan="2">18</td></tr>
<tr><td>Pillar spacing (cm)</td><td colspan="2">20</td></tr>
<tr><td>Number of trials</td><td>10</td><td>10</td></tr>
</table>

**Table S2. Sample size, mass, and dimensions of animal experiments.** Average data are mean ± s.d.

| **Body shape experiments** | **Cuboidal** | **Elliptical** |
|---|---|---|
| Number of individuals | 10 | 10 |
| Total number of trials | 99 | 117 |
| Number of trials for circular pillars | 48 | 54 |
| Number of trials for square pillars | 51 | 63 |
| Body mass (without shell) (g) | 2.4 ± 0.3 | 2.7 ± 0.2 |
| Body length (cm) | 4.9 ± 0.3 | 5.1 ± 0.2 |
| Body width (cm) | 1.9 ± 0.1 | 1.9 ± 0.1 |
| Body thickness (cm) | 0.9 ± 0.1 | 0.9 ± 0.1 |
| Shell mass (g) | 2.6 ± 0.2 | 2.7 ± 0.2 |
| Shell length (cm) | 5 | |
| Shell width (cm) | 2.7 | |
| Shell thickness (cm) | 1.2 | |
| Pillar cross section diameter/side length (cm) | 1.3 | |
| Pillar height (cm) | 7 | |
| Pillar lateral spacing (cm) | 3 | |
| **Sensor-deprivation experiments** | **Intact** | **Sensor** |
| Number of individuals | 3 | 3 |
| Number of trials | 30 | 30 |

**Table S3. List of parts and vendors.**

| Part | Specification / Material | Vendor / Manufacturer |
|---|---|---|
| Robot chassis | 0.15 cm acrylic sheet | McMaster Carr, NJ |
| Laser cutter | VLS 6.60 | Universal Laser Systems Inc, AZ |
| Servomotor | Dynamixel XL-320 | Robotis, CA |
| S-shaped legs | PLA using Ultimaker+ | Dynamism |
| Duct tape wrap for legs | - | Duck Brand, OH |
| Sandpaper for obstacle track | 60 grit size, Pro No-Slip Grip Advanced | 3M Inc, MN |
| High speed camera | Go Series | JAI Inc., CA |
| Work lamps | L14SLED, 1000 W | Designer Edge |
| Animals | - | Pinellas County Reptiles, FL |
| Vacuum former | 508FS | Formech Inc., WI |
| Pillars | Solid 6061 aluminum | McMaster Carr, NJ |
| Drone | Parrot Airborne MiniDrone | NueBlue LLC |

## Supplementary Text

### S1. Details of imaging and 3-D reconstruction

We used a small shutter time (500 μs) to reduce motion blur and a small lens aperture to maximize the focal depth of field. We illuminated the test area with four 1000 W work lamps. We calibrated the cameras for 3-D kinematics reconstruction via Direct Linear Transformation (Hedrick, 2008). The custom calibration object built from Lego bricks covered the entire field of view for maximal reconstruction precision. We verified tracking and reconstruction fidelity (s.d. of position error = 0.6 mm; s.d. of orientation error = 1.1°) using a 3-D printed high precision object. We also verified that lens distortion was minimal (<1%) using the checkboard distortion measurement method. We verified that the reconstructed 3-D motion matched with the observed motion by projecting it onto videos from each camera.

### S2. Statistics

All the probability values were calculated relative to the total number of accepted trials of each treatment. All average data are reported as means ± 1 s.d.

For both robot and animal, we performed Kruskal-Wallis tests to check statistical significance ($\alpha = 0.05$) of the following dependencies:

(1) traversal probability on obstacle shape and orientation, for each body shape;

(2) final bearing on initial bearing, for each body shape and obstacle shape and orientation treatment;

(3) maximal pitch increase on initial bearing, for each body shape and obstacle shape and orientation treatment;

(4) final bearing on body shape and obstacle shape and orientation treatment;

(5) maximal pitch increase on body shape and obstacle shape and orientation treatment;

(6) difference in final bearing between animal and robot, for each body shape and obstacle shape and orientation treatment;

(7) difference in maximal pitch increase between animal and robot, for each body shape and obstacle shape and orientation treatment;

(8) distance along trajectory travelled on body shapes in multi-pillar robot experiments (Section 6.1)

(9) difference in speed, deceleration, and absolute angular acceleration between intact and sensor-deprived animals (Section S3).

For the animal, we pooled data from all individuals and included individual as a random effect in statistical tests.

## S3. Animal's initial interaction is dominated by passive mechanics

Because cockroaches can use antennae to sense and detect obstacles to navigate complex terrain (Cowan et al., 2006; Harley et al., 2009; Okada and Toh, 2006), it is possible that obstacle interaction is dominated by sensory feedback control rather than passive mechanical interaction (Dickinson et al., 2000). To verify that this was not the case, we compared the magnitudes of speed, acceleration, and angular acceleration in the horizontal plane at initial collision for the same individuals before and after depriving them of antennae (Harley et al., 2009). None of these differed (Fig. S3; $P > 0.05$, Kruskal-Wallis test), indicating that the effect of sensory feedback control was minimal at initial contact. In addition, body pitch, yaw, and roll changed little after the antennae came in range of the obstacles before initial collision, indicating that the animal did not actively slow down further or reorient its body before collision (Fig. S2). Even after initial collision, response by the animal is delayed (~90 ms) due to delays in sensory signal transmission (40 ms (Harley et al., 2009)) and subsequent leg muscle activation (47 ms (More et al., 2010; Sponberg and Full, 2008)). Thus, the animal's initial body-obstacle interaction was mostly feedforward and dominated by passive mechanics and thus behaved similarly to the feedforward robot. It was not until the later part of interaction phase (average duration: 130 ± 110 ms) that the animal adjusted its body motion to escape obstacle attraction (Figs. 4D, 5D, S5F, S6F).

**S4. Body-obstacle interaction consists of continual collisions**

We found that body-obstacle interaction during dynamic interaction consisted of stochastic continual collisions for both animal and robot. This could be observed from body-obstacle distance, which fluctuated substantially (up to 10% body length) and irregularly about several times per second (Fig. S1A, B), with standard deviation exceeding the mean (Fig. S1C). In addition, we verified that even when interacting with a single pillar, the robot suffered large unpredicatable rotational fluctuations (Fig. S2).

**S5. Application to aerial robots**

Although discovered for terrestrial locomotion, the principles of shape-induced obstacle attraction and repulsion may also be useful for aerial or aquatic locomotion where the body physically interact with solid objects. As a proof of concept, we tested a small, off-the-shelf drone with a custom compliant cage (Fig. S8, Movie S13). When encountering a narrow spacing between two vertical pillars (110% cage width), the drone with a cuboidal cage (Fig. S8D, 10 trials) was always attracted towards the pillar it first contacted and was frequently stuck between the two pillars. By contrast, with an elliptical cage (Fig. S8B, E), it was always repelled away from the pillars into the gap and always traversed. In addition, obstacle attraction induced by the cuboidal shape enabled the drone to be perch onto a pillar (via Velcro) (Fig. S8C, F). Although drones with spherical protective cages have been demonstrated to use collisions to traverse sparse obstacles (Briod et al., 2013, 2014), our demonstration showed that desired obstacle interactions can be elicited using body shape and will be useful for aerial robots to traverse extremely cluttered environments.

**Supplementary References**

Briod A, Kornatowski P, Klaptocz A, et al. (2013) Contact-Based Navigation For An Autonomous Flying Robot. In: *Ieee/Rsj International Conference On Intelligent Robots And Systems*. Ieee, Pp. 3987–3992.

Briod A, Kornatowski P, Zufferey Jc, et al. (2014) A Collision-Resilient Flying Robot. *Journal Of Field*

*Robotics* 31: 496–509. Doi: 10.1002/Rob.21495.

Cowan NJ, Lee J And Full RJ (2006) Task-Level Control Of Rapid Wall Following In The American Cockroach. *Journal Of Experimental Biology* 209(15): 1617–1629. Doi: 10.1242/Jeb.02433.

Dickinson MH, Farley CT, Full RJ, et al. (2000) How Animals Move: An Integrative View. *Science* 288(5463): 100–106. Doi: 10.1126/Science.288.5463.100.

Harley CM, English BA and Ritzmann RE (2009) Characterization Of Obstacle Negotiation Behaviors In The Cockroach, Blaberus Discoidalis. *Journal Of Experimental Biology* 212(10): 1463–1476. Doi: 10.1242/Jeb.028381.

Hedrick TL (2008) Software Techniques For Two- And Three-Dimensional Kinematic Measurements Of Biological And Biomimetic Systems. *Bioinspiration & Biomimetics* 3(3): 034001. Doi: 10.1088/1748-3182/3/3/034001.

More HL, Hutchinson JR, Collins DF, et al. (2010) Scaling Of Sensorimotor Control In Terrestrial Mammals. *Proceedings. Biological Sciences / The Royal Society* 277(1700): 3563–8. Doi: 10.1098/Rspb.2010.0898.

Okada J and Toh Y (2006) Active Tactile Sensing For Localization Of Objects By The Cockroach Antenna. *Journal Of Comparative Physiology A: Neuroethology, Sensory, Neural, And Behavioral Physiology* 192(7): 715–726. Doi: 10.1007/S00359-006-0106-9.

Sponberg S And Full RJ (2008) Neuromechanical Response Of Musculo-Skeletal Structures In Cockroaches During Rapid Running On Rough Terrain.: 433–446. Doi: 10.1242/Jeb.012385.

**Supplementary Figures**

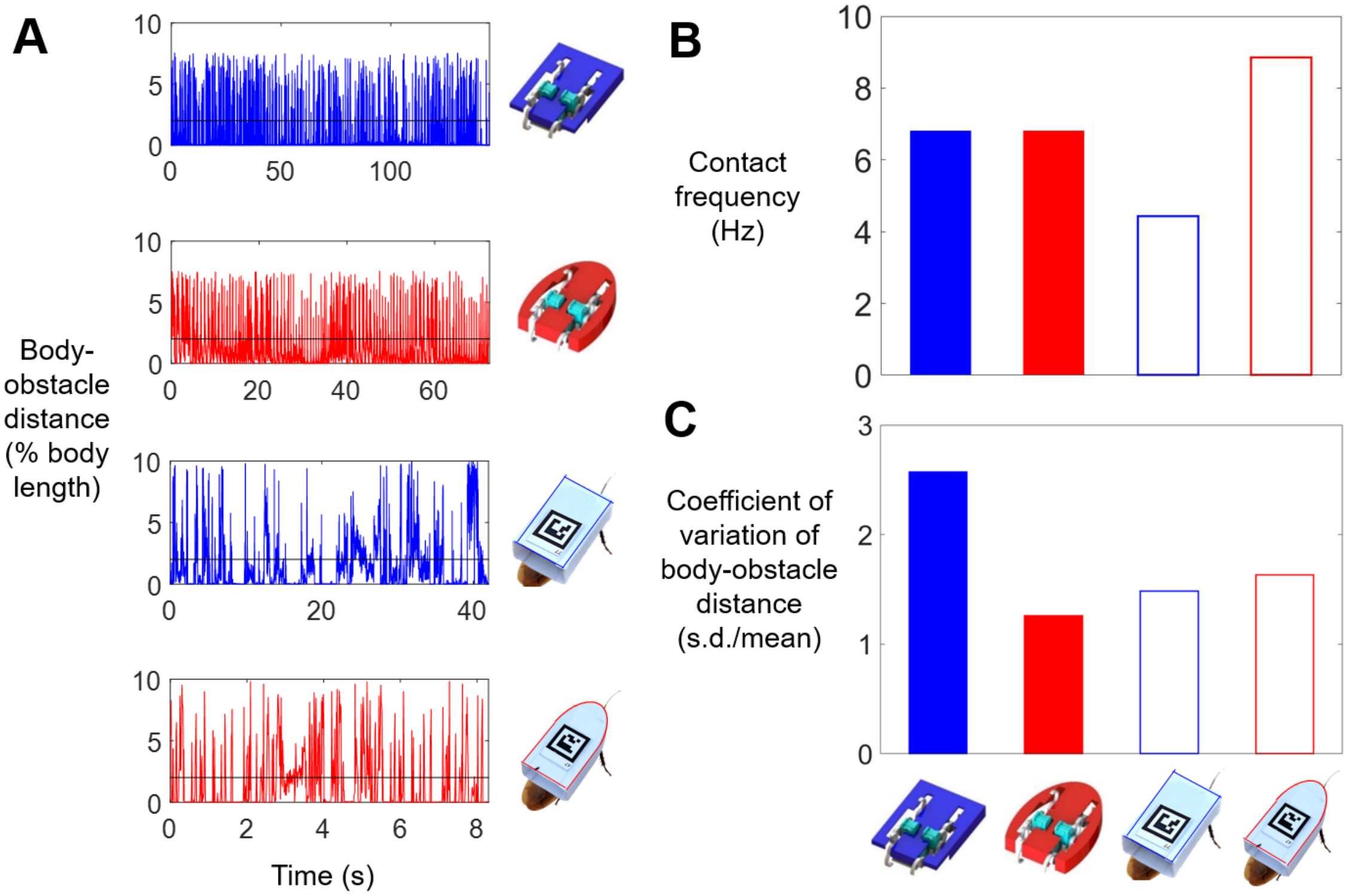

**Fig. S1. Body-obstacle interaction consists of continual collisions.** (**A**) Body-obstacle distance (in units of body length). All experimental trials are pooled onto the same time axis. Horizontal black line shows threshold of 2% body length to classify contact and non-contact. (**B**) Average contact frequency. (**C**) Coefficient of variation of body-obstacle distance (s.d./mean). Blue and red are for cuboidal and elliptical body shapes, respectively.

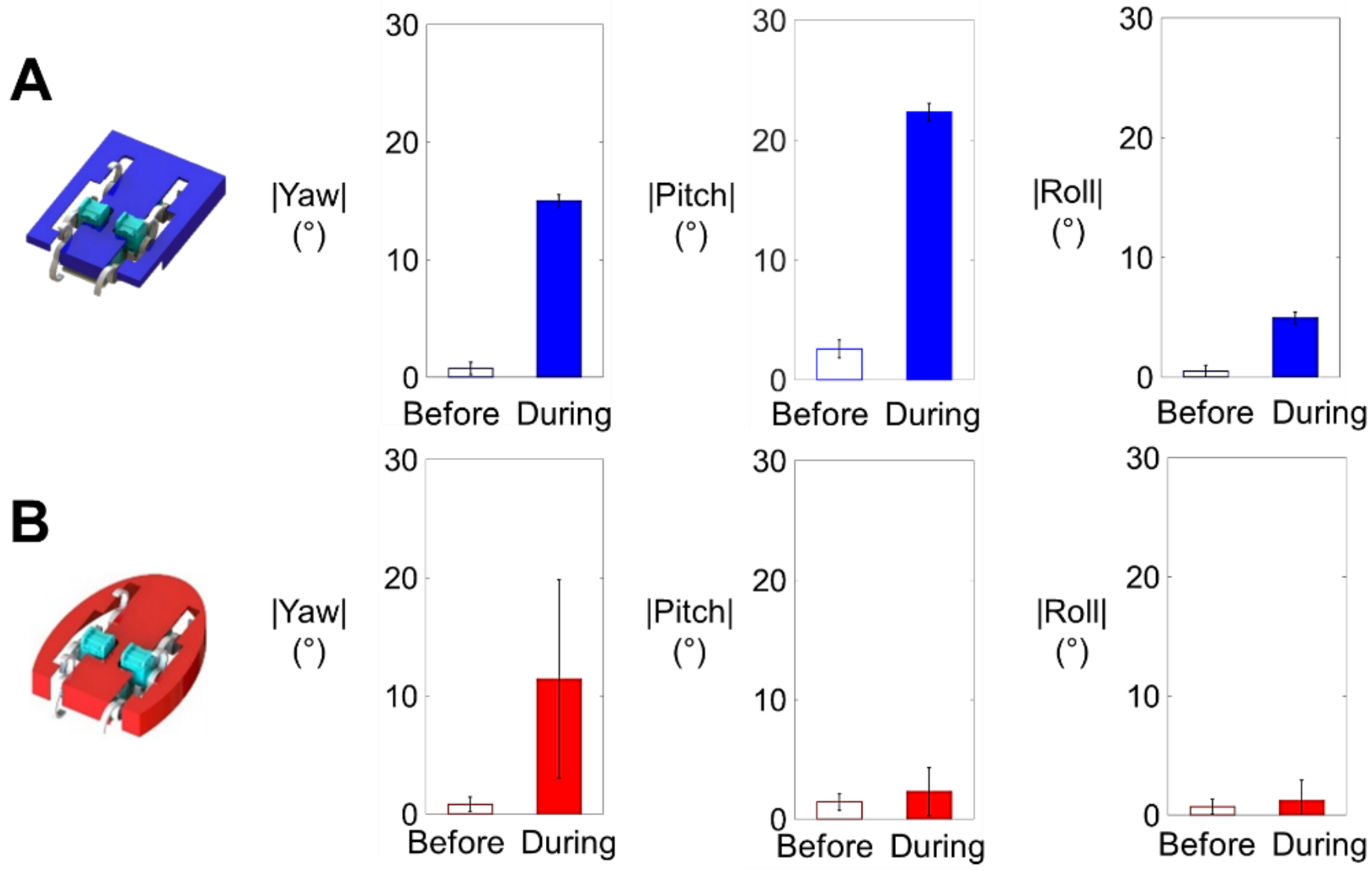

**Fig. S2. Robot orientation varies substantially during obstacle interaction.** (**A, B**) Body yaw, pitch, roll of cuboidal (A) and elliptical (B) robot in absolute values. Error bars show ± 1 s.d.

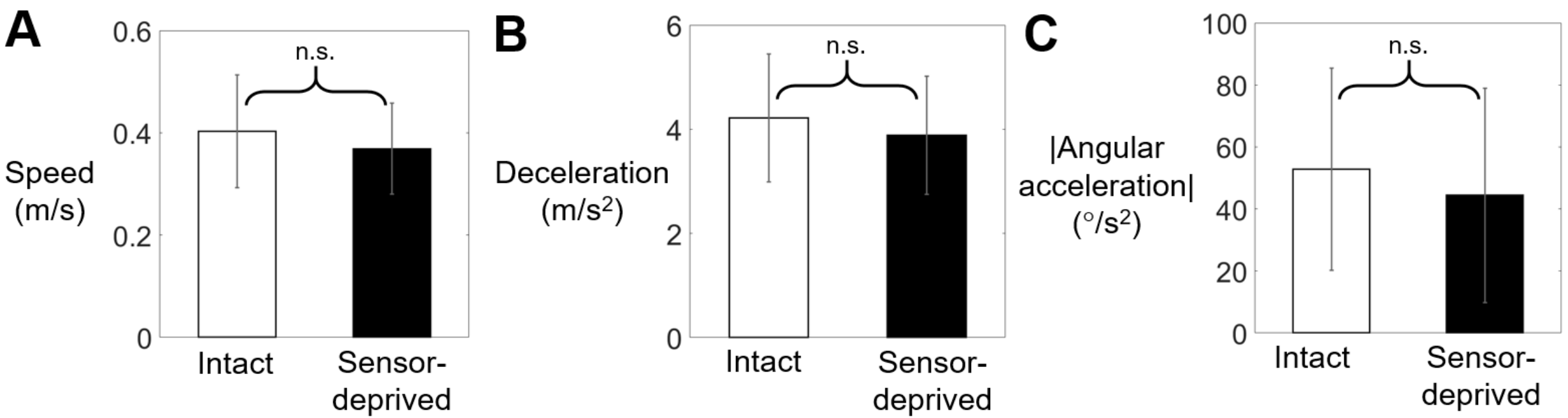

**Fig. S3. Animal's initial body-pillar collision is dominated by passive mechanics and not sensory feedback.** Comparison of (**A**) speed, (**B**) acceleration, and (**C**) angular (turning) acceleration in the horizontal (*x*-*y*) plane of the same individuals before and after removing both antennae. Error bars show ± 1 s.d. Brackets and n.s. indicate no significant difference.

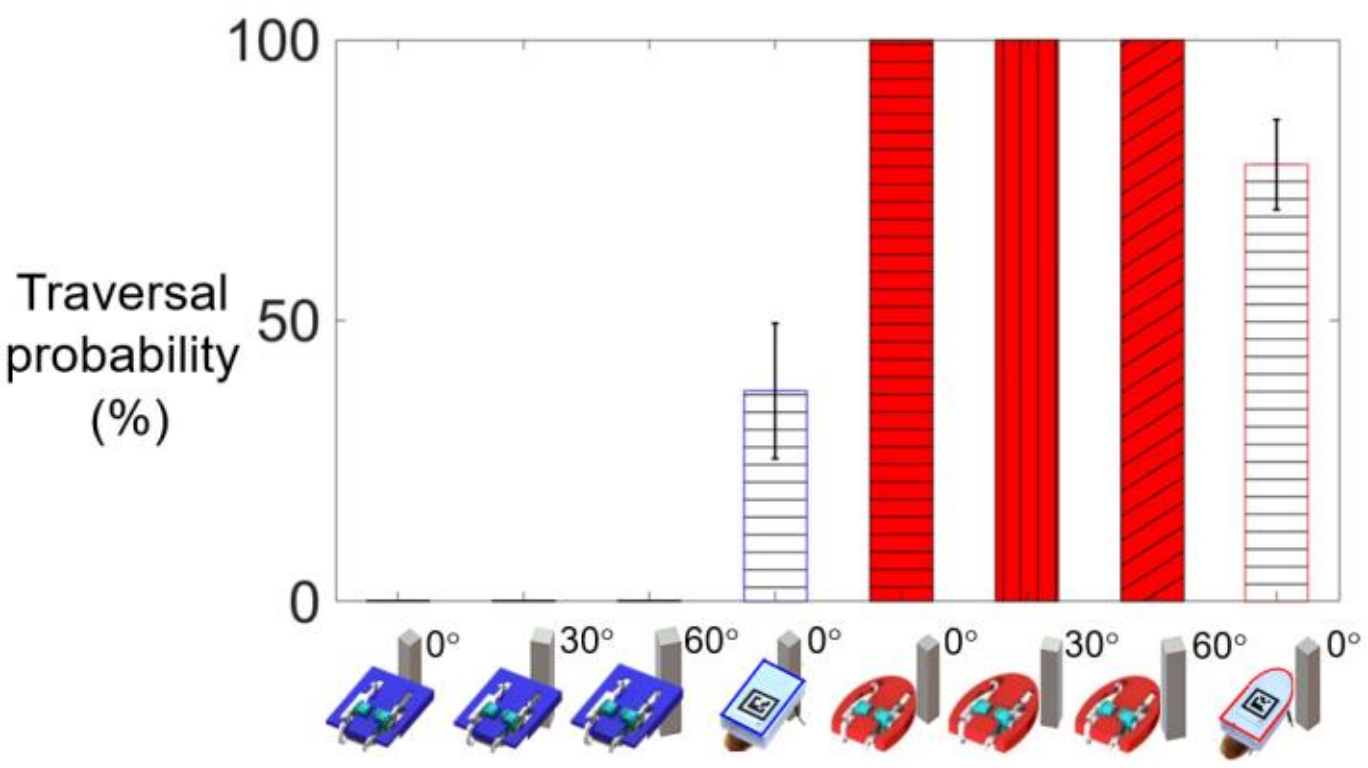

**Fig.S4. Traversal probability for square pillars.** Solid bars are for robot and empty bars are for animal. Blue is for cuboidal body and red is for elliptical body. Horizontal, vertical, and oblique hatches are for square pillars oriented at 0°, 30°, and 60° relative to $x$-axis. Error bars show ± 1 s.d. See Movie S2 for representative trials.

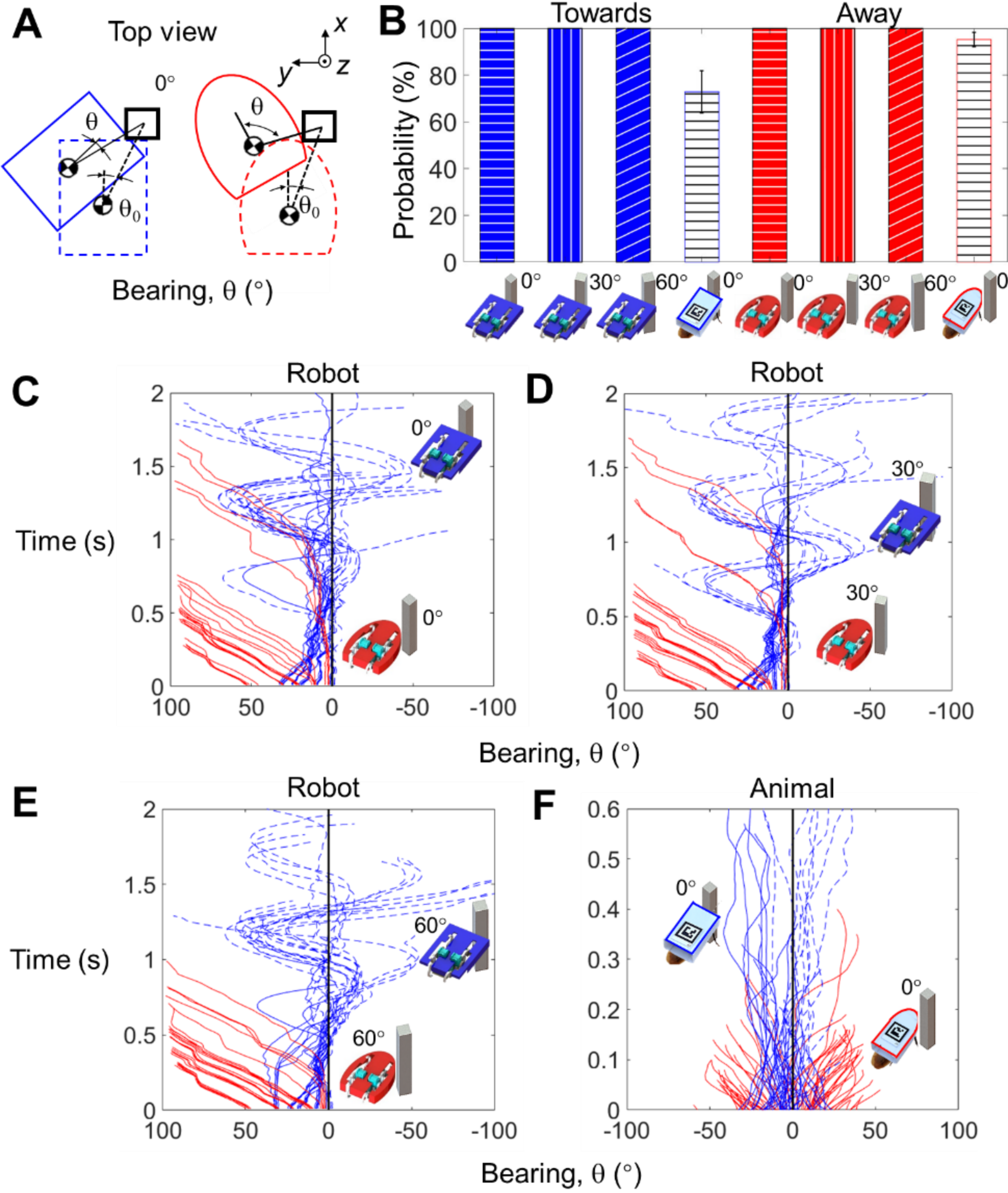

**Fig. S5. Turning motion during obstacle interaction for square pillars.** See Fig. 4 for definition of plots in (A, C-F). See Fig. S4 for definition of bar graph in (B). See Table 1 for sample size. See Movie S2 for representative trials.

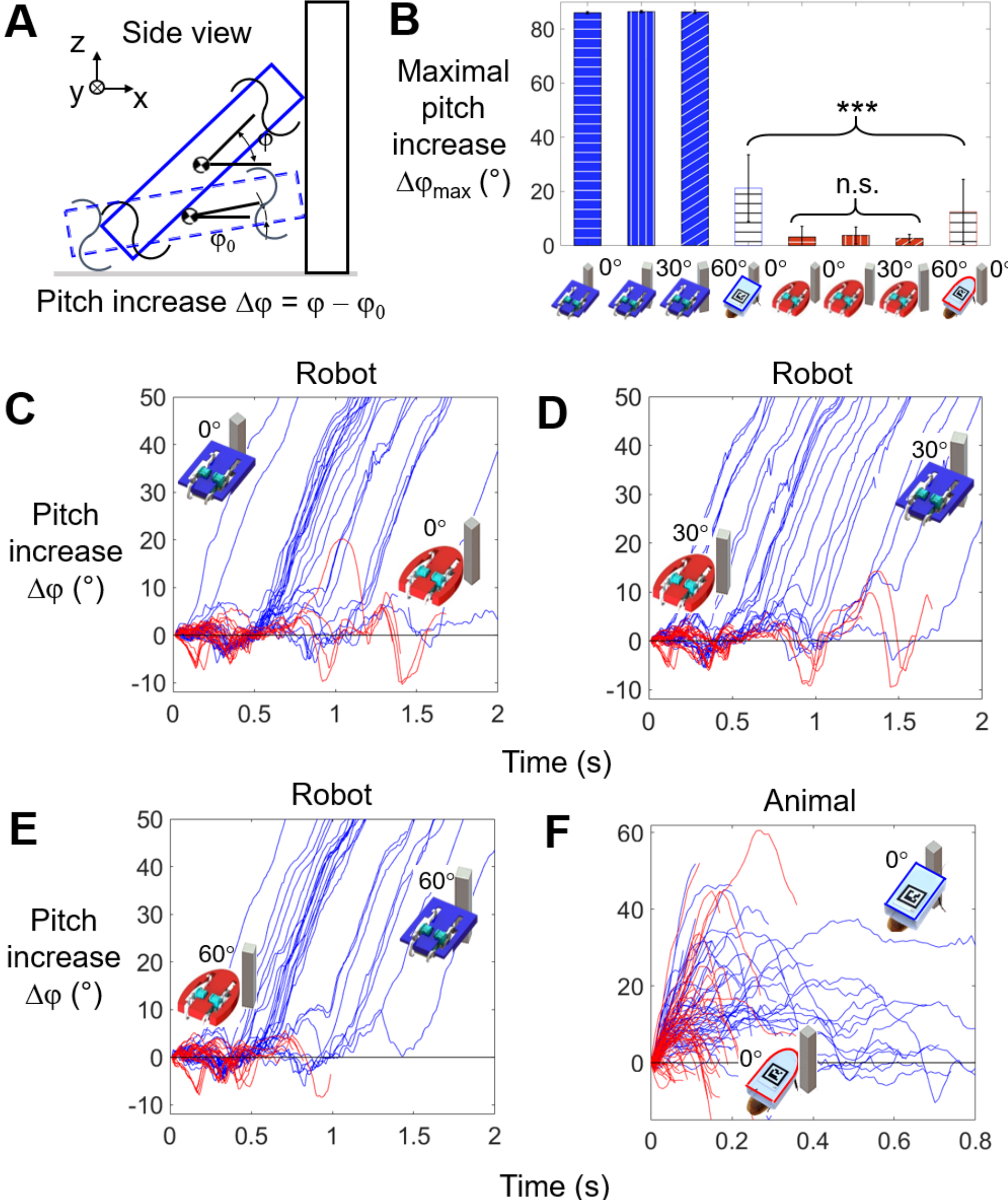

**Fig. S6. Pitching motion during obstacle interaction for square pillars.** See Fig. 5 for definition of plots in (A, C-F). See Fig. S4 for definition of bar graph in (B). See Table 1 for sample size. See Movie S2 for representative trials.

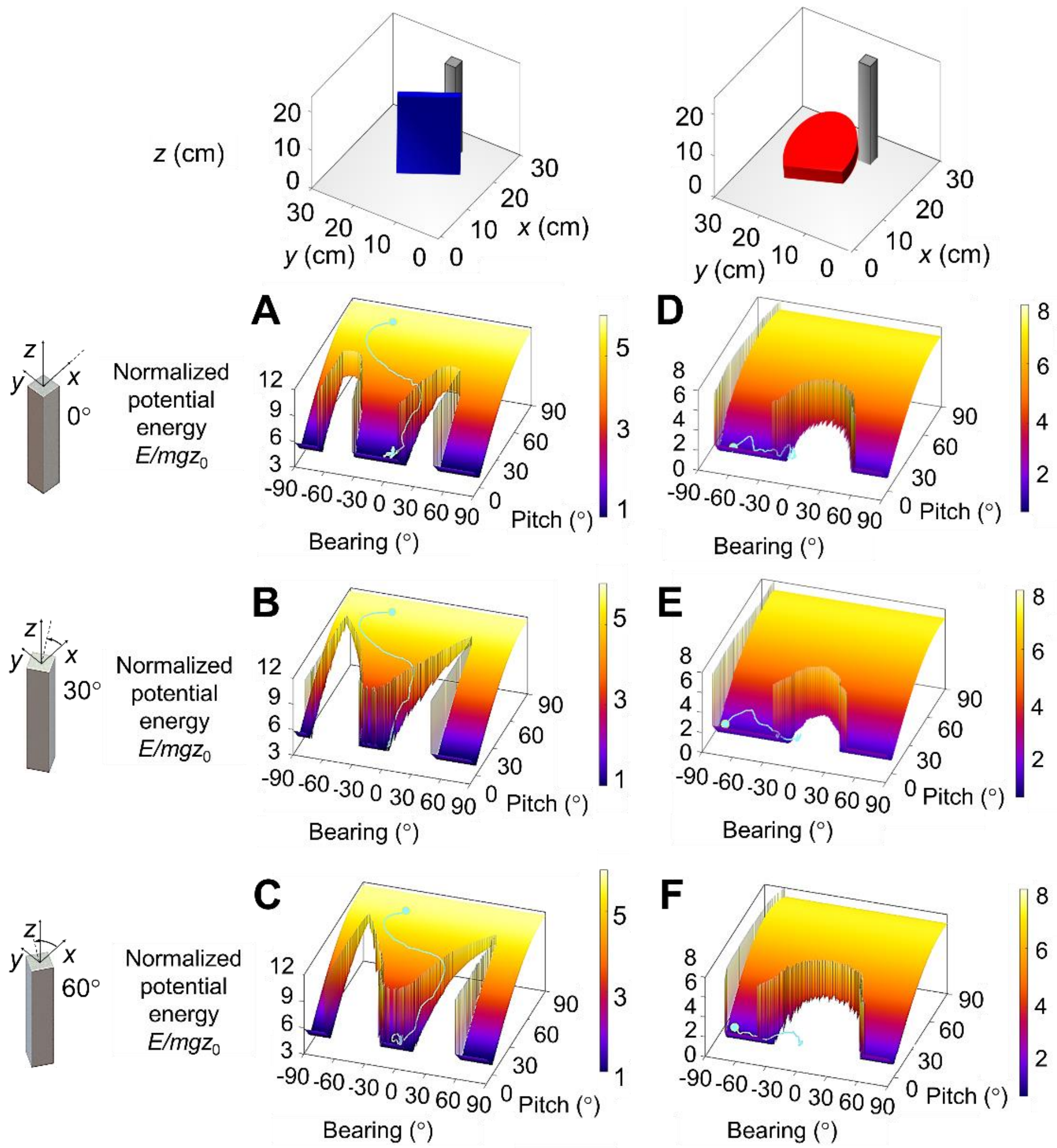

**Fig. S7. Potential energy landscape with square pillars of different orientations.** (**A**-**C**) Cuboidal and (**D-E**) elliptical robot interacting with square pillars oriented at 0° (A, D), 30° (B, E), and 60° (C, F) relative to *x*-axis. See Fig. 6 for definition of plots.

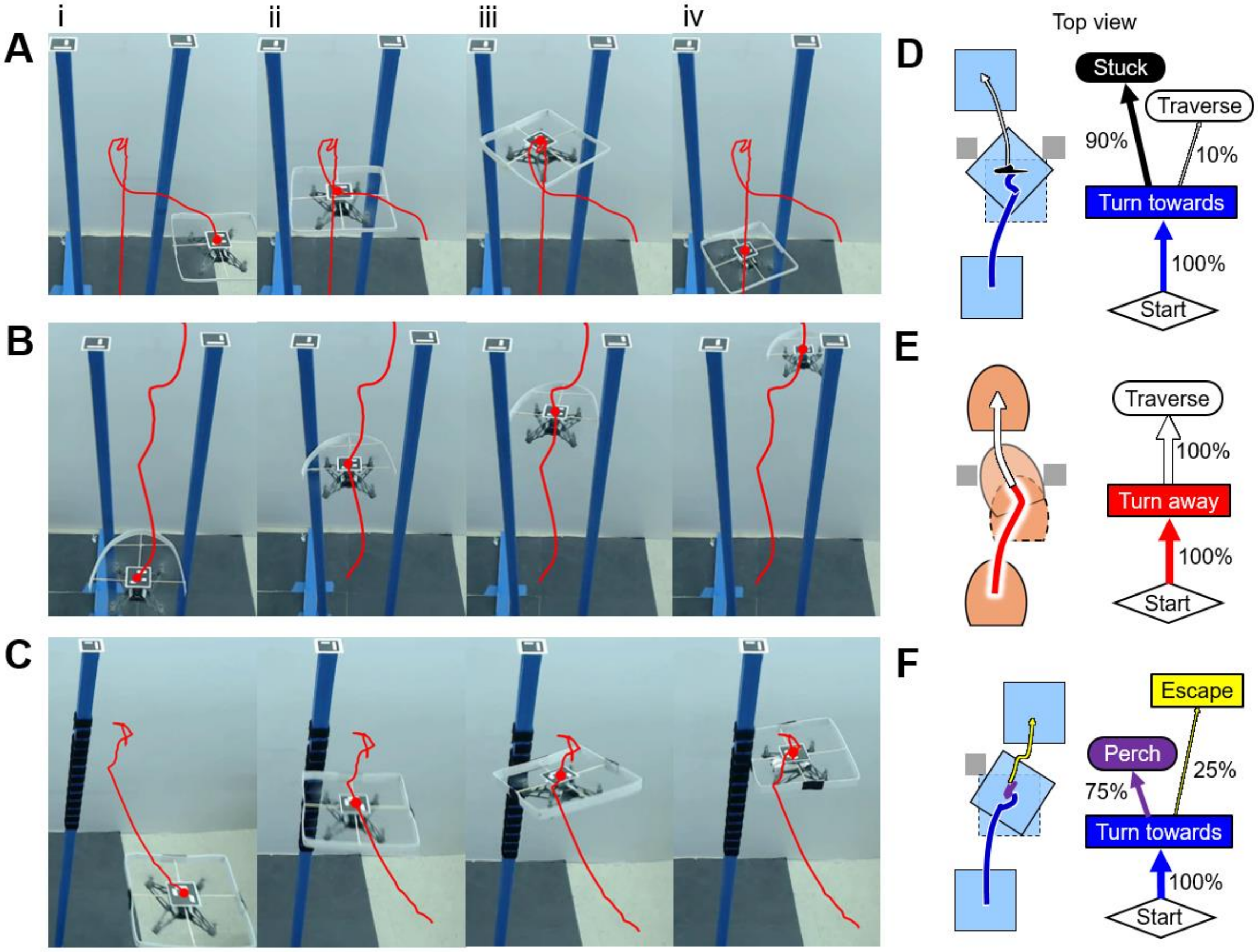

**Fig. S8. Drone using body shape-induced obstacle attraction and repulsion.** (**A**) Failure to traverse a gap between two pillars using a cuboidal body. (**B**) Gap traversal using an elliptical body. (**C**) Perching onto a pillar using a cuboidal body. Red curve shows drone trajectory. (**D, E, F)** Body turning direction (rectangular boxes) and outcome (rounded boxes) for cases shown in (A-C). Dashed body shows typical state at initial contact and solid body shows typical state resulting from interaction. Numbers next to the arrows show probability of that transition. See Table 1 for sample size.

**Supplementary Movies**

**Movie S1**. Body motion when interacting with a pillar obstacle.

**Movie S2**. Effect of body shape on interaction is insensitive to obstacle shape and orientation.

**Movie S3**. Potential energy landscape of a cuboidal body interacting with a pillar obstacle.

**Movie S4**. Potential energy landscape of an elliptical body interacting with a pillar obstacle.

**Movie S5**. Ensemble of system state trajectories on potential energy landscape with a cuboidal body.

**Movie S6**. Ensemble of system state trajectories on potential energy landscape with an elliptical body.

**Movie S7**. Potential energy landscape of intermediate body shapes between elliptical and cuboidal.

**Movie S8**. Robot moving in cluttered terrain.

**Movie S9**. Cuboidal robot moving in natural cluttered terrain using open-loop gait vs. pitch-and-turn strategy.

**Movie S10**. Robot traversing natural cluttered terrain using passive control via shape and active pitch-and-turn strategy.

**Movie S11**. Potential energy landscape predicts turning torque required for a cuboidal body to overcome that from body weight to escape obstacle attraction.

**Movie S12**. Pitching up helps a cuboidal robot turn to escape obstacle attraction.

**Movie S13**. Drone traversing a narrow obstacle gap and perching onto an obstacle using passive control via body shape.